\documentclass[a4paper,fleqn,usenatbib]{mnras}

\usepackage{newtxtext,newtxmath}

\usepackage[T1]{fontenc}
\usepackage{ae,aecompl}
\usepackage{soul}
\usepackage{ulem}

\usepackage{graphicx}	
\usepackage{amsmath}	
\usepackage{color} 

\title{Performance of different correction maps in extended phase-space method for spinning compact binaries}

\author[J. Luo et al.]{
Junjie Luo$^{1}$\thanks{luojunjie33@mail2.sysu.edu.cn}, 
Jie Feng $^{2}$\thanks{fengjie5@mail2.sysu.edu.cn},
Hong-Hao Zhang$^{1}$\thanks{zhh98@mail.sysu.edu.cn},
Weipeng Lin,$^{1}$\thanks{linweip5@mail.sysu.edu.cn}.
\\
$^{1}$ School of Physics, Sun Yat-sen University, Guangzhou 510275, China\\
$^{2}$ School of Science, Shenzhen Campus of Sun Yat-sen University, Shenzhen 518107, China\\
$^{3}$ School of Physics and Astronomy, Sun Yat-sen University, Zhuhai, China
}

\date{Accepted XXX. Received YYY; in original form ZZZ}

\pubyear{2022}

\begin{document}
\label{firstpage}
\pagerange{\pageref{firstpage}--\pageref{lastpage}}
\maketitle

\begin{abstract}
Since the first detection of gravitational waves by the LIGO/VIRGO team, the related research field has attracted more attention. 
The spinning compact binaries system, as one of
the gravitational-wave sources for broadband laser interferometers, has been widely studied by related researchers. In order to analyze the gravitational wave signals using matched filtering techniques, reliable numerical algorithms are needed. Spinning compact binaries system in Post-Newtonian (PN) celestial mechanics have inseparable Hamiltonian. The extended phase-space algorithm is an effective solution for the problem of this system. We have developed correction maps for the extended phase-space method in our previous work, which significantly improves the accuracy and stability of the method with only a momentum scale factor. In this paper we will add more scale factors to modify the numerical solution in order to minimize the errors in the constants of motion. However, we find that these correction maps will result in a large energy bias in the subterms of the Hamiltonian in chaotic orbits, whose potential and kinetic energy, etc. are calculated inaccurately. We develop new correction maps to reduce the energy bias of the subterms of the Hamiltonian, which can instead improve the accuracy of the numerical solution, and also provides a new idea for the application of the manifold correction in other algorithms.
\end{abstract}




\section{Introduction} \label{sec:intro}
The existence of gravitational waves was an important prediction after Einstein established the general theory of relativity in the early nineteenth century. Since its existence was confirmed(\cite{LIGOScientific:2016aoc}), Einstein's prediction about relativity has been fully proved by experiments. Moreover, scientists have discovered a whole new means of observing the universe, which is definitely another milestone in the development of astronomy and has attracted more attentions to the field of gravitational wave detection. The spinning compact binaries consisting of neutron stars or black holes as one of the gravitational wave sources for broadband laser interferometry is a highly nonlinear, integrable relativistic binary problem, which is a rich source of potential chaos and brings gravitational waves with stronger observable effects. The calculation of chaotic orbits of binary stars is a great challenge, and the numerical study of long-term evolution becomes very complicated and difficult. Chaos may prevent the application of matched filtering methods to extract these signals from the noise. Therefore, the successful detection of waveform should constrain chaotic parameter spaces and regions. On the other hand, the accurate calculation of chaotic orbits of spin binaries will be beneficial for broadening the detection range of gravitational waves.

For simplicity, the motion for strong gravitational systems such as spinning compact binary systems are usually described by the post-Newtonian approximation\citep{Blanchet:2002mb,Tanay2021,Zotos_2019} instead of the Einstein's equations in the situation of large distances and small velocities (much slower than the speed of light).  This makes the coordinates and momenta of the Hamiltonian become non-separable variables. Without separable coordinate and momentum forms, the Hamiltonian can't be separated into two or more non-interacting integrable parts. Therefore, explicit symplectic algorithms\citep{Feng_1987,Huang2022,Wu2021}, which are based on the operator splitting, are unavailable for spinning compact binaries in PN celestial mechanics, so implicit symplectic integrators are naturally chosen. For example, \cite{Tsang_2015APJL} developed implicit slimplectic methods for integrations of general nonconservative systems applying in a Newtonian two-body problem with 2.5PN gravitational radiation reaction terms, 2nd order of the Post-Newtonian (PN) term with the consideration of the "tail" effect in the wave zone. \citep{Lubich:2010mj} developed a 4th-order noncanonical explicit and implicit mixed symplectic integrator (using noncanonical and nonconjugate spin variables) of \cite{Suzuki:1990PLA} for a splitting approach to orbital and spin contributions. The term ``explicit" means that the spin-orbit and spin-spin Hamiltonians are solved independently and analytically, while ``implicit" means that the non-spin orbital part is computed by the implicit Euler method. Suzuki's fourth-order composition is a product of five second-order integrators.
With the construction of the canonical and conjugate spin variables \cite{Wu_2010APS}, \cite{Zhong:2010PRD} presented fourth-order canonical explicit and implicit mixed symplectic algorithms in which the second-order explicit leapfrog algorithm calculates the separable Hamiltonians and the nonseparable terms are solved by the second-order implicit midpoint method. In addition to explicit and implicit mixed symplectic integration scheme, the pure implicit integrators such as implicit midpoint method and Gauss Runge-Kutta implicit canonical symplectic schemes \cite{Seyrich:2013PRD} are also feasible.

Although the implicit algorithms are easy to be constructed in the inseparable Hamiltonian, there are inevitable problems including large consumption of computational resources by repeated iterations, and the problem of iterative divergence, which becomes more serious especially in chaotic orbits of spinning compact binaries with high nonlinearity. The extended phase-space method, which can avoid the above problems, is an alternative solution. \cite{Pihajoki:2015} extends the phase-space variables of position and momentum coordinates and presents the extended phase-space explicit methods with momenta permutation map, where the original and corresponding extended momenta exchange their values with each other at every integration step to avoid increasing differences in values over time. Based on the work of \cite{Pihajoki:2015}, \cite{Liu:2016MNRAS} developed the coordinate and momenta sequent permutation maps for the fourth-order extended phase-space explicit algorithm constructed by two Yoshida's triple products of the second-order leapfrog algorithm to have better energy error behaviour \cite{Yoshida_1990PRD}. Nevertheless, this algorithm suffers from major failures in numerical simulations of chaotic orbits, where the difference between the original and extended variables increases with time due to their interactions. Although such differences are small for regular orbits, numerically sensitive chaotic systems can amplify the differences and fall into a vicious circle. To solve this problem, we proposed a midpoint map that ensures that the original and extended variables are strictly equal, and only one Yoshida's triple product is needed to construct the fourth-order algorithm thus doubling the computational efficiency \citep{Luo:2017EPJP, Luo:2017ApJ}. In addition, \cite{Pan:2021aiu} applies the midpoint map to the coherent post-Newtonian Euler-Lagrange equations and also obtained good performance. In a recent work by \cite{Hu:2019ApJ}, the midpoint map shows excellent performances in comparison with several algorithms. The problem seems to be solved, except that the midpoint map exerted on numerical solutions might cause the total energy change. Due to those changes, some numerical simulations, such as those for chaotic orbits in spinning compact binary or restricted three-body problems, show energy error growths. With the application of the manifold correction, it is not complicated to ensure that the total energy does not change after the map exerted. For the first time, we introduced manifold corrections into the extended phase space method to improve the accuracy of the numerical solution and error stability \cite{Luo:2020}. However, in our previous work, only a single momenta scale factor is used to adjust the numerical solutions. More scale factors will be discussed in this work.

The organization of this paper is as follows. In section~\ref{sec:2}, we revisit three types of manifold corrections and design their corresponding correction maps for the extended phase-space methods, and propose a new one. In section \ref{sec:3}, we examine all correction maps in the numerical simulations of PN conservative Hamiltonian system of spinning compact binaries without the radiative terms. We use the eighth- and ninth-order Runge-Kutta-Fehlberg algorithm of variable step sizes as a reference to obtain the accuracy of numerical solutions adjusted by different correction maps. Finally, we give our conclusion in section \ref{sec:4}.

\section{Correction  map in extended phase space}\label{sec:2}
\subsection{Extended phase space method and momentum scale factor}

\begin{figure}
\centering
\includegraphics[width=0.4\textwidth]{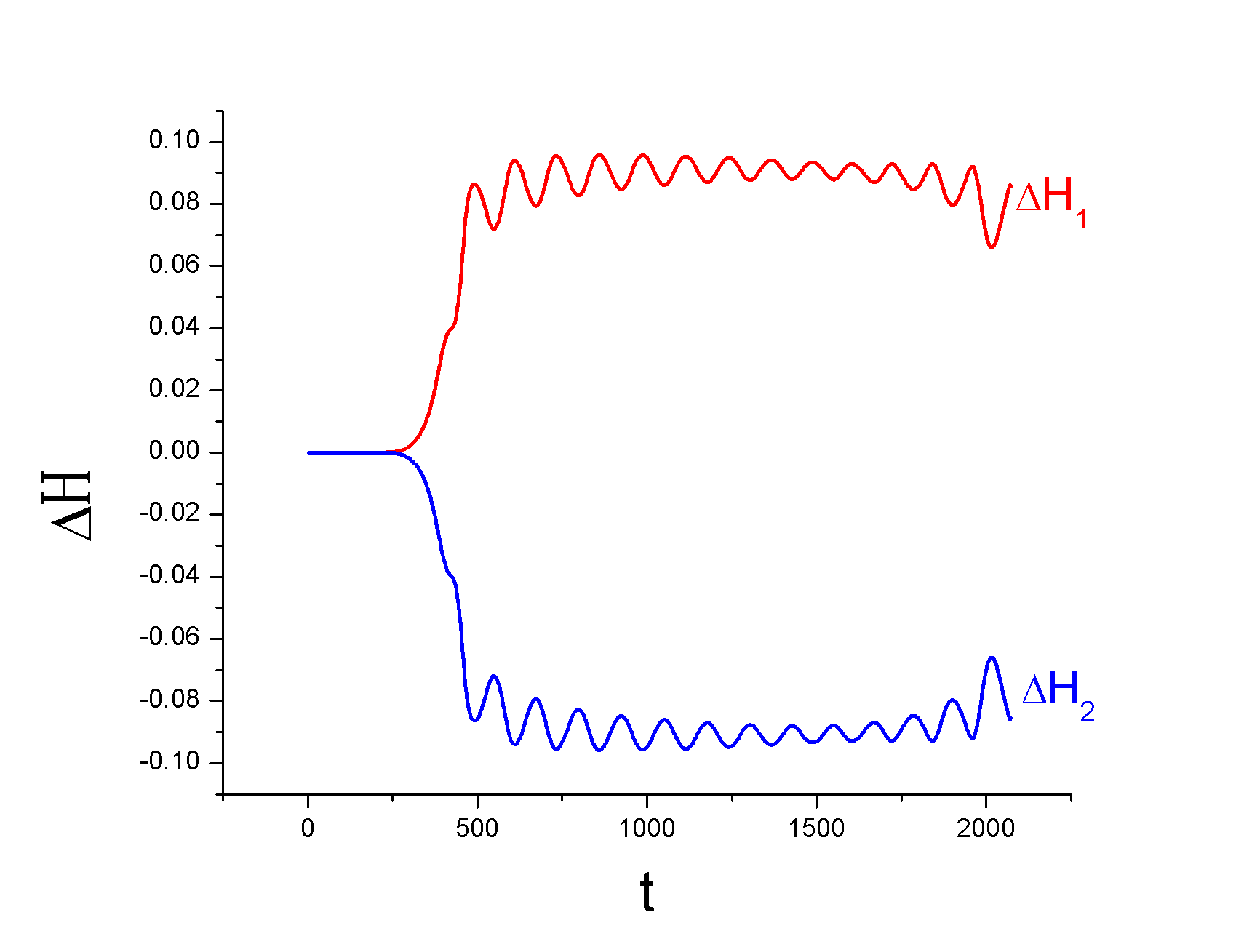}
\caption{Energy error of $H_{1}$ and $H_{2}$ calculated by extended phase-space method without any map. In this situation it is a pure explicit symmetric method for the whole Hamiltonian $\widetilde{H}$. Here the absolute energy error $\Delta \mathcal{H}=H_{i}(t)-H(0), (i=1,2)$ , and $H_{i}(t)$ correspond to the value of the Hamiltonian $H_{1}$ or $H_{2}$ at time t, and $H(0)$ is the initial value of origin Hamiltonian $H$. It is obvious that there exist symmetry between $\Delta\mathcal{H}_1$(red) and $\Delta\mathcal{H}_2$(blue).}
\label{fig1}
\end{figure}

The extended phases-space method \citet{Pihajoki:2015} is an alternative way to the implicit algorithm for the non-separable Hamiltonians, which are as functions of position $\mathbf{r}$ and momentum $\mathbf{p}$ and can not decompose into two or more integrable parts. In the extended phases-space method, the pair of canonical and conjugate variables $(\textbf{r}, \textbf{p})$ is copied to a new pair of phase-space variables ($\widetilde{\textbf{r}}$, $\widetilde{\textbf{p}}$). The two pairs of canonical and conjugate variables $(\textbf{r}, \textbf{p})$ and ($\widetilde{\textbf{r}}$, $\widetilde{\textbf{p}}$) are reorganized into new Hamiltonians in the extended phase space,
\begin{eqnarray}
       \widetilde{H}(\textbf{r},\widetilde{\textbf{r}},\textbf{p},
       \widetilde{\textbf{p}})=H_{1}(\textbf{r},
       \widetilde{\textbf{p}})+H_{2}(\widetilde{\textbf{r}},\textbf{p}).
\end{eqnarray}
Both $H_1$ and $H_2$ should be equal to the original Hamiltonian $H$. After the above processing the whole Hamiltonian $\widetilde{H}$ will contain two integrable parts, so that the standard second order leapfrog algorithm \cite{Pihajoki:2015} can be adopted as:
\begin{eqnarray}
\mathbf{A}_2(h)=\textbf{H}_{2}(\frac{h}{2})\textbf{H}_{1}(h)\textbf{H}_{2}(\frac{h}{2}),
\end{eqnarray}
Where $\textbf{H}_{1}$ and $\textbf{H}_{2}$ are Hamiltonian operators as functions of time step $h$. It is important to emphasize that the solutions $(\textbf{r}, \widetilde{\textbf{p}})$ and $(\widetilde{\textbf{r}}, \textbf{p})$ are expected to be identical at every time step. However, as time evolves, they diverge quickly due to the interplay between the solutions $(\textbf{r},\widetilde{\textbf{p}})$ of $H_{1}$ and $(\widetilde{\textbf{r}},\textbf{p})$ of $H_{2}$, as shown in Fig. \ref{fig1}, which is derived from the previous work \cite{Luo:2020}.

To find the way out, \citet{Pihajoki:2015} proposes the momentum permutation map to restrain the equality of solutions of $H_{1}$ and $H_{2}$, which fails in the chaotic orbit calculation of spinning compact binaries, until the correction map \cite{Luo:2020} is adopted,

\begin{eqnarray}
\textbf{M}_1=\left(\begin{array}{cccc}
\frac{\textbf{1}}{2},  \frac{\textbf{1}}{2},  \textbf{0},  \textbf{0} \\
\frac{\textbf{1}}{2}, \frac{\textbf{1}}{2},  \textbf{0},  \textbf{0} \\
\textbf{0},  \textbf{0}, \mathbf{\alpha}, \mathbf{\alpha} \\
\textbf{0},  \textbf{0},   \mathbf{\alpha},\mathbf{\alpha}
\end{array}\right).
\end{eqnarray}
 Here $\alpha$ is a scale factor, which can be solved by the constant of motion or in a way designed by the researcher. Then the leapfrog algorithm called the extended phase-space method with a correction map can be written as, 
\begin{eqnarray}
       \mathbf{C}^{*}_2(h)= \mathbf{A}_2(h)\textbf{M}_1=\textbf{H}_{2}(\frac{h}{2})\textbf{H}_{1}(h)\textbf{H}_{2}(\frac{h}{2})\textbf{M}_1.
\end{eqnarray}
From the $n$th to $(n+1)$th step, numerical solutions are expressed as
\begin{eqnarray}
\left(\begin{array}{cccc}
\textbf{r}\\
\widetilde{\textbf{r}}\\
\textbf{p}\\
\widetilde{\textbf{p}}
\end{array}\right)_{n+1}
=\mathbf{C}^{*}_2
\left(\begin{array}{cccc}
\textbf{r}\\
\widetilde{\textbf{r}}\\
\textbf{p}\\
\widetilde{\textbf{p}}
\end{array}\right)_{n}.
\end{eqnarray}
\cite{Liu:2016MNRAS} suggested a fourth-order explicit integrator constructed with two Yoshida's triplet products, while $\mathbf{C}^{*}_2$ consists of only one Yoshida's triplet product, which is the product of three leapfrogs $\textbf{A}_{2}$ with one correction map.
Then the fourth-order explicit extended phase-space algorithm with a correction map is set up as
\begin{eqnarray}
\textbf{C}_{4}(h)=\textbf{M}_1\otimes\textbf{A}_3(h),\label{eq:6}
\end{eqnarray}
 where $\textbf{A}_3(h)=\textbf{A}_{2}(\lambda_{3} h)\textbf{A}_{2}(\lambda_2 h)\textbf{A}_{2}(\lambda_1 h)$ and symbol $\otimes$ denotes the Kronecker product. The time coefficients $\lambda_{1},\lambda_{2}$ and $\lambda_{3}$ are in completely accord with paper \cite{Yoshida_1990PRD}. In order to achieve the fourth-order accuracy, the sum of the third-order errors of $\textbf{A}_{2}$ should be equal to zero, i.e., $\lambda_{1}^{3}+\lambda_{2}^{3}+\lambda_{3}^{3}=0$. And the sum of these time coefficients equals to one time step, i.e., $\lambda_{1}+\lambda_{2}+\lambda_{3}=1$. Here two equations are provided with three unknown parameters. For simplicity, we assume that $\lambda_{1}=\lambda_{3}$, and get the time coefficients $\lambda_{1}=\lambda_{3}=1/(2-2^{1/3})$ and $\lambda_{2}=1-2\lambda_{1}$. The correction map designed in \cite{Luo:2020} not only guarantees the equivalence between the original variables and the corresponding replicated variables, but also ensures that the value of new Hamiltonian $\widetilde{H}$ does not change after the correction map exerted, which none of the previous map schemes can do. With these advantages, the expanded phase-space method with the correction map shows excellent performance with high efficiency, stability, and high accuracy.

There is simply one scale factor in $\textbf{C}_{4}$, but the form of the correction map is not unique. Moreover, different forms of the correction map give various performance in the extended phase space method, which we will describe in detail in the next subsections.

\subsection{Different correction maps in spinning compact binaries}

In the Lagrangian formula of a spinning compact binary system, its purely orbital (non-spinning) terms can be written in the 2PN order(\cite{Blanchet:2002mb}), while the spin effects of two spinning bodies are the leading-order (1.5PN) spin-orbit coupling and the leading-order (2PN) spin-spin coupling(\cite{Hartl:2004xr}). The light speed $c$ and the constant of gravity $G$ are given in nature units with $c=G=1$. The variables evolve according to the following Lagrangian.

\begin{eqnarray}
L=L_{N}+L_{1PN}+L_{2PN}+L_{1.5SO}+L_{2SS}.
\end{eqnarray}
where 
\begin{eqnarray}
L_{N}=\frac{\mathbf{\dot{r}}^{2}}{2}+\frac{1}{r}, 
\end{eqnarray}
\begin{eqnarray}
L_{1PN} &=& \frac{1}{8}(1-3\eta)\mathbf{\dot{r}}^{4}+\frac{1}{2}[(3+\eta)\mathbf{\dot{r}}^{2}
\nonumber \\
&& +\eta(\textbf{N}\cdot\mathbf{\dot{r}})^{2}]\frac{1}{r} -\frac{1}{2r^{2}},
\end{eqnarray}

\begin{eqnarray}
L_{2PN}&=&\frac{1}{16}(1-7\eta+13\eta^{2})\mathbf{\dot{r}}^{6}+\frac{1}{8}[(7-12\eta-9\eta^{2})\mathbf{\dot{r}}^{4} \nonumber \\
&&+(4-10\eta)\eta(\textbf{N}\cdot\mathbf{\dot{r}})^{2}\mathbf{\dot{r}}^{2}
+3\eta^{2}(\textbf{N}\cdot\mathbf{\dot{r}})^{4}]\frac{1}{r} \nonumber \\
&&+\frac{1}{2}[(4-2\eta+\eta^{2})\mathbf{\dot{r}}^{2}
+3\eta(1+\eta)(\textbf{N}\cdot\mathbf{\dot{r}})^{2}]\frac{1}{r^{2}} \nonumber \\
&&+\frac{1}{4}(1+3\eta)\frac{1}{r^{3}},
\end{eqnarray}
$L_{1.5SO}$ and $L_{2SS}$ are given by \cite{Hartl:2004xr},
\begin{eqnarray}
L_{1.5SO}=-\frac{1}{r^{3}}\textbf{S}\cdot(\textbf{r}\times\dot{\mathbf{r}}),
\end{eqnarray}
\begin{eqnarray}
L_{2SS}=-\frac{1}{2r^{3}}[\frac{3}{r^{2}}(\textbf{S}_{0}\cdot\textbf{r})^{2}-\textbf{S}_{0}^{2}],
\end{eqnarray}
where $\textbf{S}= [2+3/(2\beta)]\textbf{S}_{1} +(2+3\beta/2)\textbf{S}_{2}$, $\mathbf{S}_0=(1+1/\beta)\mathbf{S}_1+(1+\beta)\mathbf{S}_2$. In the extended phase-space method, the corresponding Hamiltonian $H$ can be obtained from the Legendre transformation of the Lagrangian $L$,
\begin{eqnarray}
H=\mathbf{p}\cdot \mathbf{\dot{r}}-L,
\end{eqnarray}
\begin{eqnarray}
\mathbf{p}=\partial L/\partial \mathbf{\dot{r}}.
\end{eqnarray}
Then we get the 2PN Hamiltonian \cite{Wu_2015APS},
\begin{eqnarray}
H=H_{N}+H_{PN}+H_{SOSS}, \label{eq:19}
\end{eqnarray}

The sub-Hamiltonians in the equation \ref{eq:19} are respectively written as
 \begin{eqnarray}
H_{N}=T(\textbf{p})+V(\textbf{r})=\frac{\textbf{p}^{2}}{2}-\frac{1}{r},
\end{eqnarray}
\begin{eqnarray}
H_{PN}=H_{1PN}+H_{2PN},
\end{eqnarray}
\begin{eqnarray}
H_{1PN} &=& \frac{1}{8}(3\eta-1)\textbf{p}^{4}-\frac{1}{2}[(3+\eta)\textbf{p}^{2} \nonumber \\
&& +\eta(\textbf{N}\cdot\textbf{p})^{2}]\frac{1}{r}+\frac{1}{2r^{2}},
\end{eqnarray}
\begin{eqnarray}
H_{2PN} &=& \frac{1}{16}(1-5\eta+5\eta^{2})\textbf{p}^{6}+\frac{1}{8}[(5-20\eta-
3\eta^{2})\textbf{p}^{4} \nonumber\\ && -2\eta^{2}(\textbf{N}\cdot\textbf{p})^{2}\textbf{p}^{2}-3\eta^{2}
(\textbf{N}\cdot\textbf{p})^{4}]\frac{1}{r} \nonumber\\
&& +\frac{1}{2}[(5+8\eta)\textbf{p}^{2}+3\eta
(\textbf{N}\cdot\textbf{p})^{2}]\frac{1}{r^{2}} \nonumber\\
&& -\frac{1}{4}(1+3\eta)\frac{1}{r^{3}},
\end{eqnarray}
\begin{eqnarray}
H_{SOSS}=H_{1.5SO}+H_{2SS},
\end{eqnarray}
and
\begin{eqnarray}
H_{1.5SO}=\frac{1}{r^{3}}\textbf{S}\cdot(\textbf{r}\times\textbf{p}),
\end{eqnarray}
\begin{eqnarray}
H_{2SS}=\frac{1}{2r^{3}}[\frac{3}{r^{2}}(\textbf{S}_{0}\cdot\textbf{r})^{2}-\textbf{S}_{0}^{2}].
\end{eqnarray}

The constants of motion in this system, such as energy $E = H$, angular momenta $\mathbf{J}=\mathbf{S}_1+\mathbf{S}_2+\mathbf{r}\times \mathbf{p}$ and spin lengths $\mathbf{S}^{2}_j=S^{2}_j$, are derived from

\begin{eqnarray}
\mathbf{\dot{r}} = \frac{\partial H}{\partial\mathbf{p}}, ~~~~~~
\mathbf{\dot{p}} = -\frac{\partial H}{\partial\mathbf{r}}, ~~~~~~\mathrm{and}~~~~~~ 
\mathbf{\dot{S}}_j = \frac{\partial H}{\partial\mathbf{S}_j}\times\mathbf{S}_j.
\end{eqnarray}

The spin variables were not canonical or conjugate before the work done by \cite{Wu_2010APS}. According to the conservation of spin magnitudes, \cite{Wu_2010APS} introduce a set of generalized coordinates $\theta_j$ and generalized momenta $\xi_j$, then rewrite the unit spin vector as
\begin{eqnarray}
\hat{\textbf{S}}_{j}=
\left(\begin{array}{cccc}
\rho_{j}\cos\theta_{j}\\
\rho_{j}\sin\theta_{j}\\
\xi_{j}/S_j
\end{array}\right),
\end{eqnarray}
where $\rho_{j}=\sqrt{1-(\xi_j/S_j)^{2}}$. Above all, the Hamiltonian in Eq. \ref{eq:19} will be transformed into an equation with only canonical and conjugate phase-space variables $(\mathbf{r},\theta_{1}, \theta_{2}; \mathbf{p}, \xi_1, \xi_2)$ and can be expressed as
\begin{eqnarray}
      H(\mathbf{r},\theta_{1}, \theta_{2};\mathbf{p}, \xi_1, \xi_2)=H(\mathbf{R};\mathbf{P}). \label{eq:27}
\end{eqnarray}

Without considering the gravitational dissipation, spinning compact binaries have constants of motion such as conservation of energy and conservation of angular momentum. We assume that the initial energy is $E_0$, and the angular momentum vector $\mathbf{J}_0$ has three components $[J_{x0},J_{y0},J_{z0}]$ with a magnitude of $J_0$. With time evolution, both of the angular momentum $J$ and the Hamiltonian $H$ viewed as the energy can be given by true solution of $[\mathbf{R},\mathbf{P}]$. After doubling the number of variables for the expansion of the phase space, we should always have $H(\mathbf{R},\mathbf{P})$=$H_{1}(\mathbf{R},\widetilde{\mathbf{P}})$=$H_{2}(\widetilde{\mathbf{R}},\mathbf{P})\equiv E_0$. Nevertheless, the evolution equations in the extended phase-space scheme gives a numerical solution with various biases, i.e., the computed energy $H\neq H_1\neq H_2\neq E_0$, computed angular momentum $\mathbf{J}=\mathbf{S}_1+\mathbf{S}_2+\mathbf{r}\times \mathbf{p}\neq\mathbf{J}_0$, and the computed spin length $|\hat{S}_i|\neq 1$. What's more, these biases grow as the computational procedure continues. Can the spatial scale transformations constrain the computed solution on the proper integral surfaces, so that the solution becomes a good approximation to the true solution? Several correction methods will be discussed to answer this question.

$\emph{Method 1}$: The single scale factor map for complete consistency of initial energy,

\begin{eqnarray}
\textbf{M}_1=\left(\begin{array}{cccc}
\frac{\textbf{1}}{2},  \frac{\textbf{1}}{2},  \textbf{0},  \textbf{0} \\
\frac{\textbf{1}}{2}, \frac{\textbf{1}}{2},  \textbf{0},  \textbf{0} \\
\textbf{0},  \textbf{0}, \mathbf{\alpha}, \mathbf{\alpha} \\
\textbf{0},  \textbf{0},  \mathbf{\alpha}, \mathbf{\alpha}
\end{array}\right).
\end{eqnarray}
This method looks similar to $\textbf{C}_{4}$. Instead of using the midpoint energy of $H_1$ and $H_2$ to solve for the scale factor $\alpha$ in $\textbf{C}_{4}$,
\begin{eqnarray}
       H(\frac{\textbf{R}+\widetilde{\textbf{R}}}{2},\alpha(\textbf{P}+\widetilde{\textbf{P}})) =\frac{H_{1}(\textbf{R},\widetilde{\textbf{P}})+H_{2}(\widetilde{\textbf{R}},\textbf{P})}{2}, \label{eq:9}
\end{eqnarray}
we will work out $\alpha$ with the following formula,
\begin{eqnarray}
H(\frac{\mathbf{R}+\widetilde{\mathbf{R}}}{2},\alpha(\mathbf{P}+\widetilde{\mathbf{P}})) =E_0. \label{eq:E0}
\end{eqnarray}
To distinguish it from $\textbf{C}_{4}$, we will abbreviate $\emph{Method 1}$ as $\textbf{CM}_{1}$,
\begin{eqnarray}
\textbf{CM1}(h)=\textbf{M}_1\otimes\textbf{A}_3(h).
\end{eqnarray}
$\textbf{A}_3$ is the symplectic algorithm for $\widetilde{H}$, so $\textbf{C}_{4}$ can effectively suppress the energy drift without changing the value of $\widetilde{H}$ but cannot guarantee that $\widetilde{H}=2E_0$. While, $\textbf{CM1}$ assures that $\widetilde{H}$ is equal to twice of the initial energy, i.e., $\widetilde{H}=2H=2E_0$. In order to label the numerical solutions before and after corrections, we use ($\textbf{r}^{*}, \widetilde{\textbf{r}}^{*}, \mathbf{\theta}^*_{j}, \widetilde{\mathbf{\theta}}^*_{j}; \textbf{p}^{*}, \widetilde{\textbf{p}}^{*}, \mathbf{\xi}^*_{j}, \widetilde{\mathbf{\xi}}^*_{j}$) to represent the corrected solutions, and its relationship with computed solutions ($\textbf{r}, \widetilde{\textbf{r}}, \mathbf{\theta}_{j}, \widetilde{\mathbf{\theta}}_{j}; \textbf{p}, \widetilde{\textbf{p}}, \mathbf{\xi}_{j}, \widetilde{\mathbf{\xi}}_{j}$) of $\textbf{A}_3$ is
$(\textbf{r}^{*}, \widetilde{\textbf{r}}^{*}, \mathbf{\theta}^*_{j}, \widetilde{\mathbf{\theta}}^*_{j}; \textbf{p}^{*}, \widetilde{\textbf{p}}^{*}, \mathbf{\xi}^*_{j}, \widetilde{\mathbf{\xi}}^*_{j})=$
$(\frac{(\mathbf{r}+\widetilde{\textbf{r}})}{2},\frac{(\mathbf{r}+\widetilde{\textbf{r}})}{2},\frac{(\mathbf{\theta}_j+\widetilde{\theta}_j)}{2},\frac{(\mathbf{\theta}_j+\widetilde{\theta}_j)}{2};\alpha (\mathbf{p}+\widetilde{\textbf{p}}),\alpha (\mathbf{p}+\widetilde{\textbf{p}}),\mathbf{\alpha}(\mathbf{\xi}_j+\widetilde{\mathbf{\xi}}_j),\mathbf{\alpha}(\mathbf{\xi}_j+\widetilde{\mathbf{\xi}}_j))$. 

$\emph{Method 2}$: The double scale factor map for respective consistency of the total energy and the total angular momentum,
\begin{eqnarray}
\textbf{M}_2=\left(\begin{array}{cccccccc}
 \mathbf{\frac{\gamma}{2}},   \mathbf{\frac{\gamma}{2}},  \textbf{0},  \textbf{0},  \textbf{0},  \textbf{0},  \textbf{0},  \textbf{0} \\
 \mathbf{\frac{\gamma}{2}}, \mathbf{\frac{\gamma}{2}},  \textbf{0},  \textbf{0},  \textbf{0},  \textbf{0},  \textbf{0},  \textbf{0} \\
\textbf{0},  \textbf{0}, \mathbf{\frac{1}{2}}, \mathbf{\frac{1}{2}},  \textbf{0},  \textbf{0},  \textbf{0},  \textbf{0} \\
\textbf{0},  \textbf{0},  \mathbf{\frac{1}{2}},\mathbf{\frac{1}{2}},  \textbf{0},  \textbf{0},  \textbf{0},  \textbf{0} \\
\textbf{0},  \textbf{0},  \textbf{0},  \textbf{0},  \frac{\mathbf{\alpha}}{2}, \frac{\mathbf{\alpha}}{2},  \textbf{0},  \textbf{0} \\
\textbf{0},  \textbf{0},  \textbf{0},  \textbf{0},  \frac{\mathbf{\alpha}}{2}, \frac{\mathbf{\alpha}}{2},  \textbf{0},  \textbf{0} \\
\textbf{0},  \textbf{0},  \textbf{0},  \textbf{0},  \textbf{0},  \textbf{0}, \frac{\textbf{1}}{2},  \frac{\textbf{1}}{2} \\
\textbf{0},  \textbf{0},  \textbf{0},  \textbf{0},  \textbf{0},  \textbf{0}, \frac{\textbf{1}}{2}, \frac{\textbf{1}}{2}
\end{array}\right).\label{M2}
\end{eqnarray}
With the new map, formula \ref{eq:6} will be replaced by the following one,
\begin{eqnarray}
\textbf{CM2}(h)=\textbf{M}_2\otimes\textbf{A}_3(h).
\end{eqnarray}
Its corresponding $n$th to $(n+1)$th transition is also changed to,
\begin{eqnarray}
\left(\begin{array}{cccc}
\textbf{r}\\
\widetilde{\textbf{r}}\\
\mathbf{\theta}_j\\
\widetilde{\mathbf{\theta}_j}\\
\textbf{p}\\
\widetilde{\textbf{p}}\\
\mathbf{\xi}_j\\
\widetilde{\mathbf{\xi}_j}\\
\end{array}\right)_{n+1}
=\mathbf{CM2}
\left(\begin{array}{cccc}
\textbf{r}\\
\widetilde{\textbf{r}}\\
\mathbf{\theta}_j\\
\widetilde{\mathbf{\theta}_j}\\
\textbf{p}\\
\widetilde{\textbf{p}}\\
\mathbf{\xi}_j\\
\widetilde{\mathbf{\xi}_J}\\
\end{array}\right)_{n}.
\end{eqnarray}

Unlike $\textbf{M}_1$, $\textbf{M}_2$ has two scale factors $\alpha$ and $\gamma$ adjusting the computed momenta and positions,
\begin{eqnarray}
\textbf{r}^*=\widetilde{\textbf{r}}^*=\frac{\gamma}{2}(\textbf{r}+\widetilde{\textbf{r}})\nonumber\\
\textbf{p}^*=\widetilde{\textbf{p}}^*=\frac{\alpha}{2}(\textbf{p}+\widetilde{\textbf{p}}),
\end{eqnarray}
$\alpha$ and $\gamma$ are driven by the following equations,
\begin{eqnarray}
H(\frac{\gamma(\textbf{r}+\widetilde{\textbf{r}})}{2},\frac{\alpha(\textbf{p}+\widetilde{\textbf{p}})}{2}) =E_0, \label{eq:se}
\end{eqnarray}
\begin{eqnarray}
|\alpha \gamma\textbf{L}+\textbf{S}_1+\textbf{S}_2|=J_0 \label{eq:sj}
\end{eqnarray}

$\emph{Method 3}$: The triple scale factor map with the complete consistency of the initial spin length and least-squares correction of the total energy and the magnitude of the total angular momentum.
\begin{eqnarray}
\textbf{M}_3=\left(\begin{array}{cccccccc}
\mathbf{\frac{\gamma}{2}},   \mathbf{\frac{\gamma}{2}},  \textbf{0},  \textbf{0},  \textbf{0},  \textbf{0},  \textbf{0},  \textbf{0} \\
 \mathbf{\frac{\gamma}{2}}, \mathbf{\frac{\gamma}{2}},  \textbf{0},  \textbf{0},  \textbf{0},  \textbf{0},  \textbf{0},  \textbf{0} \\
\textbf{0},  \textbf{0}, \mathbf{\frac{\delta_{j1}}{2}}, \mathbf{\frac{\delta_{j1}}{2}},  \textbf{0},  \textbf{0},  \textbf{0},  \textbf{0} \\
\textbf{0},  \textbf{0},  \mathbf{\frac{\delta_{j1}}{2}},\mathbf{\frac{\delta_{j1}}{2}},  \textbf{0},  \textbf{0},  \textbf{0},  \textbf{0} \\
\textbf{0},  \textbf{0},  \textbf{0},  \textbf{0},  \frac{\mathbf{\alpha}}{2}, \frac{\mathbf{\alpha}}{2},  \textbf{0},  \textbf{0} \\
\textbf{0},  \textbf{0},  \textbf{0},  \textbf{0},  \frac{\mathbf{\alpha}}{2}, \frac{\mathbf{\alpha}}{2},  \textbf{0},  \textbf{0} \\
\textbf{0},  \textbf{0},  \textbf{0},  \textbf{0},  \textbf{0},  \textbf{0},  \mathbf{\frac{\delta_{j2}}{2}},  \mathbf{\frac{\delta_{j2}}{2}} \\
\textbf{0},  \textbf{0},  \textbf{0},  \textbf{0},  \textbf{0},  \textbf{0},  \mathbf{\frac{\delta_{j2}}{2}},  \mathbf{\frac{\delta_{j2}}{2}}
\end{array}\right).\label{M3}
\end{eqnarray}
The scale factors $\mathbf{\delta_{j1}}$ and $\mathbf{\delta_{j2}}$ are used to keep the length of spin equal to 1, i.e.,
\begin{eqnarray}
|\hat{\textbf{S}}_{j}(\delta_{j1}\theta_j,\delta_{j2}\xi_j)|=1, j=1,2. \label{eq:cs}
\end{eqnarray}
A simple treatment that implements Eq.\ref{eq:cs} is\\
\begin{eqnarray}
\hat{\textbf{S}}_{j}(\delta_{j1}\theta_j,\delta_{j2}\xi_j)    & \ = \ &   
 \hat{\textbf{S}}_{j}(\theta_j,\xi_j)/ |\hat{\textbf{S}}_{j}(\theta_j,\xi_j)|\nonumber\\
 \left(\begin{array}{cccc}
\rho_{j}\cos(\delta_{j1}\theta_{j})\\
\rho_{j}\sin(\delta_{j1}\theta_{j})\\
\delta_{j2}\xi_{j}/S_j
\end{array}\right)    & \ = \ &
\left(\begin{array}{cccc}
\rho_{j}\cos\theta_{j}/|\hat{\textbf{S}}_{j}|\\
\rho_{j}\sin\theta_{j}/|\hat{\textbf{S}}_{j}|\\
\xi_{j}/S_j/ |\hat{\textbf{S}}_{j}|
\end{array}\right).\label{eq:cs2}
\end{eqnarray}

Therefore, $\mathbf{\delta_{j1}}$ and $\mathbf{\delta_{j2}}$ can be obtained from the Eq.\ref{eq:cs2}. Then $\gamma$ and $\alpha$ satisfy the boundary conditions
\begin{eqnarray}
\frac{\partial}{\partial \alpha}\psi(\alpha,\gamma)=0, \\ \nonumber
\frac{\partial}{\partial \gamma}\psi(\alpha,\gamma)=0,
\end{eqnarray}
where

\begin{align}
&\psi(\alpha,\gamma)=\nonumber\\
&w_1[H(\frac{\alpha (\mathbf{r}+\widetilde{\textbf{r}})}{2},\frac{\gamma (\mathbf{p}+\widetilde{\textbf{p}})}{2},\frac{\mathbf{\delta}_{j1}(\mathbf{\theta}_j+\widetilde{\theta}_j)}{2},\frac{\mathbf{\delta}_{j2}(\mathbf{\xi}_j+\widetilde{\mathbf{\xi}}_j)}{2})-E_{0}]^2\nonumber +\\
&w_2[J(\frac{\alpha (\mathbf{r}+\widetilde{\textbf{r}})}{2},\frac{\gamma (\mathbf{p}+\widetilde{\textbf{p}})}{2},\frac{\delta_{j1}(\theta_j+\widetilde{\theta}_j)}{2},\frac{\delta_{j2}(\xi_j+\widetilde{\xi}_j)}{2})-J_{0}]^2.
\end{align}

Here $w1$ and $w2$ are positive weight coefficients. The formula for the length of the spin vector is quite simple, so the solutions of Eq.$\ref{eq:cs2}$ does not consume many computational resources.
The fourth-order extended phase-space method with $\textbf{M}_{3}$ is referred to $\textbf{CM}_{3}$. It is worth mentioning that the $\textbf{C}_{4}$ ensures the equality between $H(\textbf{R}, \widetilde{\textbf{P}})$ after applying the map and $(H_{1}+H_{2})/2=\widetilde{H}(\textbf{R},\widetilde{\textbf{R}},\textbf{P}, \widetilde{\textbf{P}})/2$ before the map exerted at each integration step.
While $\textbf{CM1}$, $\textbf{CM2}$ and $\textbf{CM3}$ force the integrated solution back to the original integral hypersurface in different correction paths. $\textbf{CM1}$ focuses on improving the energy accuracy. The only difference between $\textbf{CM1}$ and $\textbf{C}_{4}$ is the calculation of $\alpha$. In $\textbf{CM}_{2}$, the number of scale factors ($\alpha,\gamma$) is two, which equals to the number of integrals. So $\textbf{CM2}$ aims to keep the energy and the angular momentum of the system being constant. $\textbf{CM3}$ considers the correction of spin vectors, in addition to the conservation of energy and angular momentum.  Unlike the Newton's method in $\textbf{CM2}$, one needs to use the least-squares method to get the optimal scaling values in $\textbf{CM3}$.

$\emph{Method 4}$:The triple scale factor map to reduce the biases of subterms of the Hamiltonian,
\begin{eqnarray}
\textbf{M}_{4}=\left(\begin{array}{cccccccc}
 \mathbf{\frac{\gamma}{2}},   \mathbf{\frac{\gamma}{2}},  \textbf{0},  \textbf{0},  \textbf{0},  \textbf{0},  \textbf{0},  \textbf{0} \\
 \mathbf{\frac{\gamma}{2}}, \mathbf{\frac{\gamma}{2}},  \textbf{0},  \textbf{0},  \textbf{0},  \textbf{0},  \textbf{0},  \textbf{0} \\
\textbf{0},  \textbf{0}, \mathbf{\frac{\alpha}{2}}, \mathbf{\frac{\alpha}{2}},  \textbf{0},  \textbf{0},  \textbf{0},  \textbf{0} \\
\textbf{0},  \textbf{0},  \mathbf{\frac{\alpha}{2}},\mathbf{\frac{\alpha}{2}},  \textbf{0},  \textbf{0},  \textbf{0},  \textbf{0} \\
\textbf{0},  \textbf{0},  \textbf{0},  \textbf{0}, \frac{\textbf{1}}{2},  \frac{\textbf{1}}{2},  \textbf{0},  \textbf{0} \\
\textbf{0},  \textbf{0},  \textbf{0},  \textbf{0},  \frac{\textbf{1}}{2},  \frac{\textbf{1}}{2},  \textbf{0},  \textbf{0} \\
\textbf{0},  \textbf{0},  \textbf{0},  \textbf{0},  \textbf{0},  \textbf{0},  \mathbf{\frac{\delta}{2}},  \mathbf{\frac{\delta}{2}} \\
\textbf{0},  \textbf{0},  \textbf{0},  \textbf{0},  \textbf{0},  \textbf{0},  \mathbf{\frac{\delta}{2}},  \mathbf{\frac{\delta}{2}}
\end{array}\right).\label{M4}
\end{eqnarray}
$\emph{Method 4}$, abbreviated as $\textbf{CM4}$,  has a momentum scale factor $\gamma$, a coordinate scale factor $\alpha$, and a spin scale factor $\delta$, which are solved by the following three equations,
\begin{align}
       T(\frac{\alpha(\textbf{p}+\widetilde{\textbf{p}})}{2})=\frac{\widetilde{T}(\textbf{p},\widetilde{\textbf{p}})}{2} =\frac{T_{1}(\widetilde{\textbf{p}})+T_{2}(\textbf{p})}{2}, \label{eq:HT}
\end{align}
\begin{align}
      & V(\frac{\gamma(\textbf{r}+\widetilde{\textbf{r}})}{2})+H_{PN}(\frac{\gamma(\textbf{r}+\widetilde{\textbf{r}})}{2},\frac{\alpha(\textbf{p}+\widetilde{\textbf{p}})}{2}) \nonumber\\
      & =\frac{\widetilde{V}(\textbf{r},\widetilde{\textbf{r}})+\widetilde{H}_{PN}(\textbf{r},\widetilde{\textbf{r}},\textbf{p},\widetilde{\textbf{p}})}{2}, \label{eq:HPN}
\end{align}
\begin{align}
      & H_{SOSS}(\frac{\gamma(\textbf{r}+\widetilde{\textbf{r}})}{2},\frac{\alpha(\textbf{p}+\widetilde{\textbf{p}})}{2},\frac{\mathbf{\theta}_{j}+\widetilde{\mathbf{\theta}}_{j}}{2},\frac{\delta(\mathbf{\xi}_{j}+\widetilde{\mathbf{\xi}}_{j})}{2}) \nonumber\\
      &=\frac{\widetilde{H}_{SOSS}}{2}(\textbf{R},\widetilde{\textbf{R}},\textbf{P},\widetilde{\textbf{P}}). \label{eq:HSOSS}
\end{align}
One can easily solve Eq.\ref{eq:HT} to get $\alpha=\sqrt{\frac{2(\textbf{p}^2+\widetilde{\textbf{p}}^2)}{(\textbf{p}+\widetilde{\textbf{p}})^2}}$. After knowing the value of $\alpha$, $\gamma$ can be obtained from Eq.\ref{eq:HPN} with the Newton's method. Then, Eq.\ref{eq:HSOSS} contains only one unknown factor $\delta$. Finally, the corresponding fourth-order extended phase space algorithm can be written as
\begin{eqnarray}
\textbf{CM4}(h)=\textbf{M}_{4}\otimes\textbf{A}_3(h).
\end{eqnarray}
 From the $n$th to $(n+1)$th step, the numerical solutions are expressed as
\begin{eqnarray}
\left(\begin{array}{cccc}
\textbf{r}\\
\widetilde{\textbf{r}}\\
\mathbf{\theta}_j\\
\widetilde{\mathbf{\theta}}_j\\
\textbf{p}\\
\widetilde{\textbf{p}}\\
\mathbf{\xi}_j\\
\widetilde{\mathbf{\xi}}_J
\end{array}\right)_{n+1}
=\mathbf{CM4}
\left(\begin{array}{cccc}
\textbf{r}\\
\widetilde{\textbf{r}}\\
\mathbf{\theta}_j\\
\widetilde{\mathbf{\theta}}_j\\
\textbf{p}\\
\widetilde{\textbf{p}}\\
\mathbf{\xi}_j\\
\widetilde{\mathbf{\xi}}_J
\end{array}\right)_{n}.
\end{eqnarray}
The numerical expression from step $n$th to $(n+1)$th is as follows
\begin{align}
&\textbf{R}_{n+\frac{1}{6}}=\textbf{R}_{n}+\frac{\lambda_1h}{2}\nabla_{\textbf{P}}H_{2}(\widetilde{\textbf{R}}_{n},\textbf{P}_{n})\nonumber\\
&\widetilde{\textbf{P}}_{n+\frac{1}{6}}=\widetilde{\textbf{P}}_{n}-\frac{\lambda_1h}{2}\nabla_{\widetilde{\textbf{R}}}H_{2}(\widetilde{\textbf{R}}_{n},\textbf{P}_{n})\nonumber\\
&\widetilde{\textbf{R}}_{n+\frac{2}{6}}=\widetilde{\textbf{R}}_{n}+\lambda_1h\nabla_{\widetilde{\textbf{P}}}H_{1}(\textbf{R}_{n+\frac{1}{6}},\widetilde{\textbf{P}}_{n+\frac{1}{6}})\nonumber\\
&\textbf{P}_{n+\frac{2}{6}}=\textbf{P}_{n}-\lambda_1h\nabla_{\textbf{R}}H_{1}(\textbf{R}_{n+\frac{1}{6}},\widetilde{\textbf{P}}_{n+\frac{1}{6}})\nonumber\\
&\widetilde{\textbf{P}}_{n+\frac{2}{6}}=\widetilde{\textbf{P}}_{n+\frac{1}{6}}-\frac{\lambda_1h}{2}\nabla_{\widetilde{\textbf{R}}}H_{2}(\widetilde{\textbf{R}}_{n+\frac{2}{6}},\textbf{P}_{n+\frac{2}{6}})\nonumber\\
&\textbf{R}_{n+\frac{2}{6}}=\textbf{R}_{n+\frac{1}{6}}+\frac{\lambda_1h}{2}\nabla_{\textbf{P}}H_{2}(\widetilde{\textbf{R}}_{n+\frac{2}{6}},\textbf{P}_{n+\frac{2}{6}})\nonumber\\
&\textbf{R}_{n+\frac{3}{6}}=\textbf{R}_{n+\frac{2}{6}}+\frac{\lambda_2h}{2}\nabla_{\textbf{P}}H_{2}(\widetilde{\textbf{R}}_{n+\frac{2}{6}},\textbf{P}_{n+\frac{2}{6}})\nonumber\\
&\widetilde{\textbf{P}}_{n+\frac{3}{6}}=\widetilde{\textbf{P}}_{n+\frac{2}{6}}-\frac{\lambda_2h}{2}\nabla_{\widetilde{\textbf{R}}}H_{2}(\widetilde{\textbf{R}}_{n+\frac{2}{6}},\textbf{P}_{n+\frac{2}{6}})\nonumber\\
&\widetilde{\textbf{R}}_{n+\frac{4}{6}}=\widetilde{\textbf{R}}_{n+\frac{2}{6}}+\lambda_2h\nabla_{\widetilde{\textbf{P}}}H_{1}(\textbf{R}_{n+\frac{3}{6}},\widetilde{\textbf{P}}_{n+\frac{3}{6}})\nonumber\\
&\textbf{P}_{n+\frac{4}{6}}=\textbf{P}_{n+\frac{2}{6}}-\lambda_2h\nabla_{\textbf{R}}H_{1}(\textbf{R}_{n+\frac{3}{6}},\widetilde{\textbf{P}}_{n+\frac{3}{6}})\nonumber\\
&\widetilde{\textbf{P}}_{n+\frac{4}{6}}=\widetilde{\textbf{P}}_{n+\frac{3}{6}}-\frac{\lambda_2h}{2}\nabla_{\widetilde{\textbf{R}}}H_{2}(\widetilde{\textbf{R}}_{n+\frac{4}{6}},\textbf{P}_{n+\frac{4}{6}})\nonumber\\
&\textbf{R}_{n+\frac{4}{6}}=\textbf{R}_{n+\frac{5}{6}}+\frac{\lambda_2h}{2}\nabla_{\textbf{P}}H_{2}(\widetilde{\textbf{R}}_{n+\frac{4}{6}},\textbf{P}_{n+\frac{4}{6}})\nonumber\\
&\textbf{R}_{n+\frac{5}{6}}=\textbf{R}_{n+\frac{4}{6}}+\frac{\lambda_3h}{2}\nabla_{\textbf{P}}H_{2}(\widetilde{\textbf{R}}_{n+\frac{4}{6}},\textbf{P}_{n+\frac{4}{6}})\nonumber\\
&\widetilde{\textbf{P}}_{n+\frac{5}{6}}=\widetilde{\textbf{P}}_{n+\frac{4}{6}}-\frac{\lambda_3h}{2}\nabla_{\widetilde{\textbf{R}}}H_{2}(\widetilde{\textbf{R}}_{n+\frac{4}{6}},\textbf{P}_{n+\frac{4}{6}})\nonumber\\
&\widetilde{\textbf{R}}_{n+1}=\widetilde{\textbf{R}}_{n+\frac{4}{6}}+\lambda_3h\nabla_{\widetilde{\textbf{P}}}H_{1}(\textbf{R}_{n+\frac{5}{6}},\widetilde{\textbf{P}}_{n+\frac{5}{6}})\nonumber\\
&\textbf{P}_{n+1}=\textbf{P}_{n+\frac{4}{6}}-\lambda_3h\nabla_{\textbf{R}}H_{1}(\textbf{R}_{n+\frac{5}{6}},\widetilde{\textbf{P}}_{n+\frac{5}{6}})\nonumber\\
&\widetilde{\textbf{P}}_{n+1}=\widetilde{\textbf{P}}_{n+\frac{5}{6}}-\frac{\lambda_3h}{2}\nabla_{\widetilde{\textbf{R}}}H_{2}(\widetilde{\textbf{R}}_{n+1},\textbf{P}_{n+1})\nonumber\\
&\textbf{R}_{n+1}=\textbf{R}_{n+\frac{5}{6}}+\frac{\lambda_3h}{2}\nabla_{\textbf{P}}H_{2}(\widetilde{\textbf{R}}_{n+1},\textbf{P}_{n+1})\nonumber\\
&\alpha=\sqrt{\frac{2(\textbf{p}^2+\widetilde{\textbf{p}}^2)}{(\textbf{p}+\widetilde{\textbf{p}})^2}}\nonumber\\
&\gamma=solve[Eq.\ref{eq:HPN},\gamma]\nonumber\\
&\delta=solve[Eq.\ref{eq:HSOSS},\delta]\nonumber
\end{align}
\begin{align}
&\textbf{r}=\frac{\gamma(\textbf{r}_{n+1}+\widetilde{\textbf{r}}_{n+1})}{2},\textbf{r}_{n+1}=\widetilde{\textbf{r}}_{n+1}=\textbf{r}\nonumber\\
&\mathbf{\theta}_{j}=\frac{(\mathbf{\theta}_{j(n+1)}+\widetilde{\mathbf{\theta}}_{j(n+1)})}{2},\mathbf{\theta}_{j(n+1)}=\widetilde{\mathbf{\theta}}_{j(n+1)}=\mathbf{\theta}_{j}\nonumber\\
&\textbf{p}=\frac{\alpha(\textbf{p}_{n+1}+\widetilde{\textbf{p}}_{n+1})}{2},\textbf{p}_{n+1}=\widetilde{\textbf{p}}_{n+1}=\textbf{p}\nonumber\\
&\mathbf{\xi}_{j}=\frac{\delta(\mathbf{\xi}_{j(n+1)}+\widetilde{\mathbf{\xi}}_{j(n+1)})}{2},\mathbf{\xi}_{j(n+1)}=\widetilde{\mathbf{\xi}}_{j(n+1)}=\mathbf{\xi}_{j}. \label{eq:10}
\end{align}

The final solutions are given by 

\begin{eqnarray}
\left(\begin{array}{cccc}
\textbf{r}^{*}\\
\widetilde{\textbf{r}}^{*}\\
\mathbf{\theta}^*_j\\
\widetilde{\mathbf{\theta}}^*_j\\
\textbf{p}^*\\
\widetilde{\textbf{p}}^*\\
\mathbf{\xi}^*_j\\
\widetilde{\mathbf{\xi}}^*_J
\end{array}\right)
=
\left(\begin{array}{cccc}
\frac{\gamma (\mathbf{r}+\widetilde{\textbf{r}})}{2}\\
\frac{\gamma (\mathbf{r}+\widetilde{\textbf{r}})}{2}\\
\frac{(\mathbf{\theta}_j+\widetilde{\theta}_j)}{2}\\
\frac{(\mathbf{\theta}_j+\widetilde{\theta}_j)}{2}\\
\frac{\alpha (\mathbf{p}+\widetilde{\textbf{p}})}{2}\\
\frac{\alpha (\mathbf{p}+\widetilde{\textbf{p}})}{2}\\
\frac{\mathbf{\delta}(\mathbf{\xi}_j+\widetilde{\mathbf{\xi}}_j)}{2}\\
\frac{\mathbf{\delta}(\mathbf{\xi}_j+\widetilde{\mathbf{\xi}}_j)}{2}
\end{array}\right).
\end{eqnarray}

There are three functions in $\textbf{CM4}$. Firstly, it ensures that $H_1$ is equal to $H_2$ to avoid the difference in energy which leads to the unavailability of the numerical solution. Secondly, it ensures that the value of $\widetilde{H}$ does not change after correction, thus suppressing the growth of the energy error. Thirdly, it reduces the energy deviation of the subterms of $H$ from half of the corresponding subterms of $\widetilde{H}$ after correction.

\begin{table*}
\begin{center}
\small \caption{Basic characteristics of correction methods in a double-precision environment.}
\label{Table 1}
\begin{tabular}{cccccc}\hline
methods & $\textbf{C}_4$ & $\textbf{CM}_1$ & $\textbf{CM}_2$ & $\textbf{CM}_3$ & $\textbf{CM}_4$ \\
\hline
 Corrected variables & $\textbf{r}$ & $\textbf{r}$  & $(\textbf{r},\textbf{p})$ & $(\textbf{r},\textbf{p},\mathbf{\theta}_j,\mathbf{\xi}_j)$ & $(\textbf{r},\textbf{p},\mathbf{\theta}_j,\mathbf{\xi}_j)$ \\
 Energy error & unknown & $ \approx10^{-16}$  & $\approx10^{-16}$ & $\approx10^{-16}$ & unknown\\
Criteria & $E=\frac{H_1+H_2}{2}$ & $E=E_0$  & $E=E_0, J=J_0$ & $E=E_0, J=J_0, |S_j|=1 $ & $T=\frac{T_1+T_2}{2}, V+H_{PN}=\frac{\widetilde{V}+\widetilde{H}_{PN}}{2}, H_{SOSS}=\frac{\widetilde{H}_{SOSS}}{2}$  \\
\hline
\end{tabular}
\end{center}
\end{table*}
Each of these four algorithms has its own characteristics. In summary, we list some features of these algorithms in Table \ref{Table 1}. For further understanding of the correction effects of each algorithm, numerical simulations are presented in the next section.

\section{Numerical simulations}\label{sec:3}

We are mainly interested in the performance of these algorithms. The methods introduced in section $\ref{sec:2}$ will be applied to control the numerical errors of PN systems of spinning compact binaries with Hamiltonian formulation $\ref{eq:19}$. There are four integrals of motion (the total energy and three integrals of the total angular momentum vector) in a ten-dimensional phase space of the canonical spin Hamiltonian. However, the absence of a fifth integral leads to the nonintegrability of this system. As a result, chaos may occur in some spin Hamiltonians (\cite{Zhong:2010PRD}; \cite{Mei_2013EPJC, Mei:2013uqa}; \cite{Luo:2020}). Now we consider a chaotic orbit, called orbit 1, in the numerical simulations, whose initial conditions are $(\beta;\textbf{r},\textbf{p})=(1;7.5,0,0,0,0.52,0), \chi_{1}=\chi_{2}=1, \hat{\textbf{S}}_{1}=(\rho_{1}\cos\frac{\upi}{4},\rho_{1}\sin\frac{\pi}{4},-0.983734), \hat{\textbf{S}}_{2}=(\rho_{2}\cos\frac{\pi}{4},\rho_{2}\sin\frac{\pi}{4},-0.983734), \rho_{1}=\rho_{2}=\sqrt{1-(-0.983734)^{2}}$, where the mass ratio $\beta=m_1/m_2$ ($m_1\leq m_2$), the total mass $m=m_1+m_2$, the reduced mass $\mu=m_1m_2/m$, the dimensionless mass parameter $\eta=\mu/m$, and $\mathbf{S}_{j}=S_{j}\mathbf{\hat{S}}_{j}$ ($j=1,2$). 
Here $\mathbf{\hat{S}}_{j}$ are unit spin vectors, and spin magnitudes $S_{j}=\chi_{j}m_{j}^{2}/m^{2}$($0\leq\chi_{j}\leq1$). The positive weight coefficients in $\textbf{CM3}$ are set to be $w1$=200 and $w2$=1. Then expanding the phase space according to the procedure presented in Section \ref{sec:2}, we obtain the new Hamiltonian $\widetilde{H}$, so that $\textbf{C}_4$, $\textbf{CM1}$, $\textbf{CM2}$, $\textbf{CM3}$ and $\textbf{CM4}$ will be available in the calculation of $\widetilde{H}$. As a reference solution, an eighth- and ninth-order Runge-Kutta-Fehlberg algorithm of variable step sizes $8(9)RKF$ will also be used to calculate the Eq. \ref{eq:19}.

\begin{figure}
\centering
\includegraphics[width=0.4\textwidth]{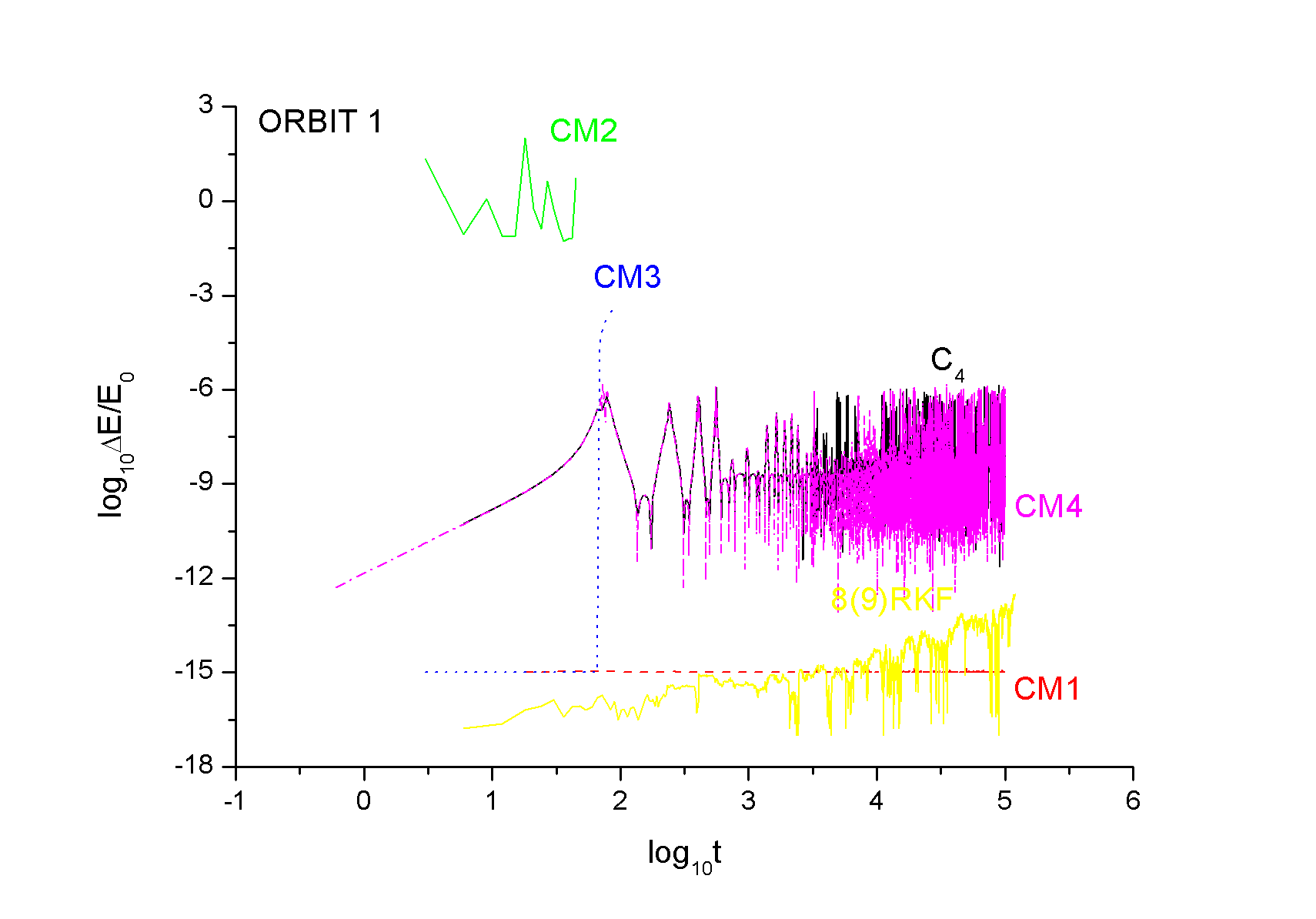}
\caption{Relative Energy error of $H$, $\Delta E/E_0=|\frac{H(t)-H(0)}{H(0)}|$, where $H(t)$ is the value of the Hamiltonian $H$ at time t, while $H(0)$ represents the initial energy. $8(9)RKF$ (yellow) has the highest accuracy, but its error increases with time steps. It is expected that $\textbf{CM1}$ (red) has excellent error performance as a method of energy-accurate correction. The energy calculated by $\textbf{CM2}$ (green) appears to have the most biased values. $\textbf{CM3}$ (blue) has high accuracy at the beginning, but quickly gets stuck in the iterative divergence. Both $\textbf{C}_4$ (black) and $\textbf{CM4}$ (purple) show compatible high accuracy and long-term stability.}
\label{fig2}
\end{figure}
Fig. \ref{fig2} shows relative energy errors with a fixed step size $h=0.6$. Among all methods, $8(9)RKF$ has the highest accuracy, but its errors increases with time steps. It is expected that $\textbf{CM1}$ has excellent error performance as a method of energy-accurate correction. The energy calculated by $\textbf{CM2}$ appears to have the most biased values. $\textbf{CM3}$ has high accuracy at the beginning, but quickly gets stuck in the iterative divergence. Both $\textbf{C}_4$ and $\textbf{CM4}$ show compatible high accuracy and long-term stability. 
\begin{figure}
\centering
\includegraphics[width=0.4\textwidth]{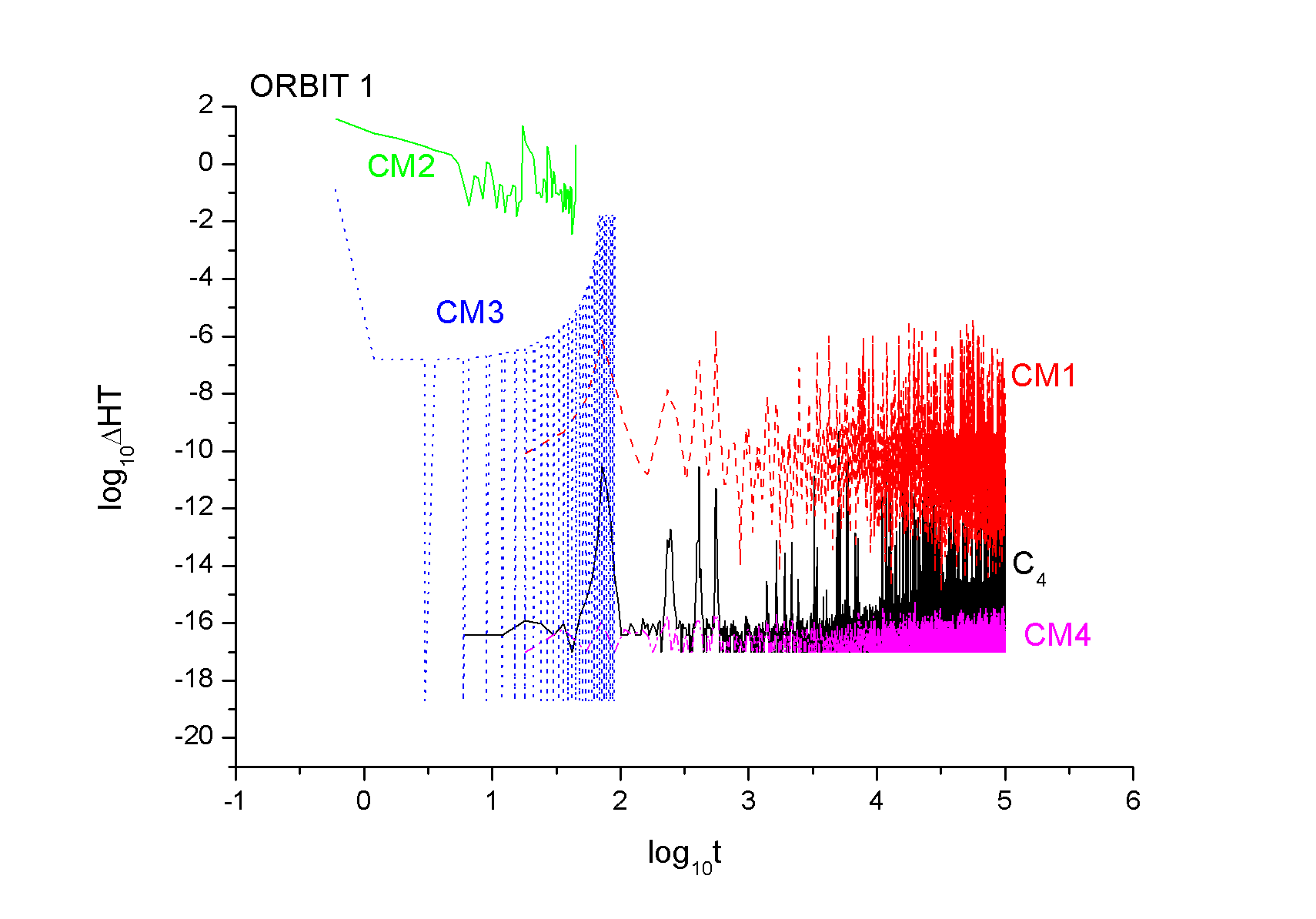}
\caption{Error behaviors of kinetic energy $T$, $\Delta HT=|T(\mathbf{p}^*)-\frac{\widetilde{T}(\mathbf{p},\widetilde{\mathbf{p}})}{2}|$. Here $\mathbf{p}^*$ denotes the momentum after corrections, while $\mathbf{p}$ before.
$\textbf{CM4}$ (purple) has the minimum bias in the kinetic energy term, while $\textbf{C}_4$ (black) has larger bias, which grows slowly. $\textbf{CM1}$ (red) shows stable error but significantly larger than $\textbf{C}_4$ and $\textbf{CM4}$. $\textbf{CM2}$ (green) and $\textbf{CM3}$ (blue) show much larger errors and cause interruptions during the calculations.
}
\label{fig3}
\end{figure}
\begin{figure}
\centering
\includegraphics[width=0.4\textwidth]{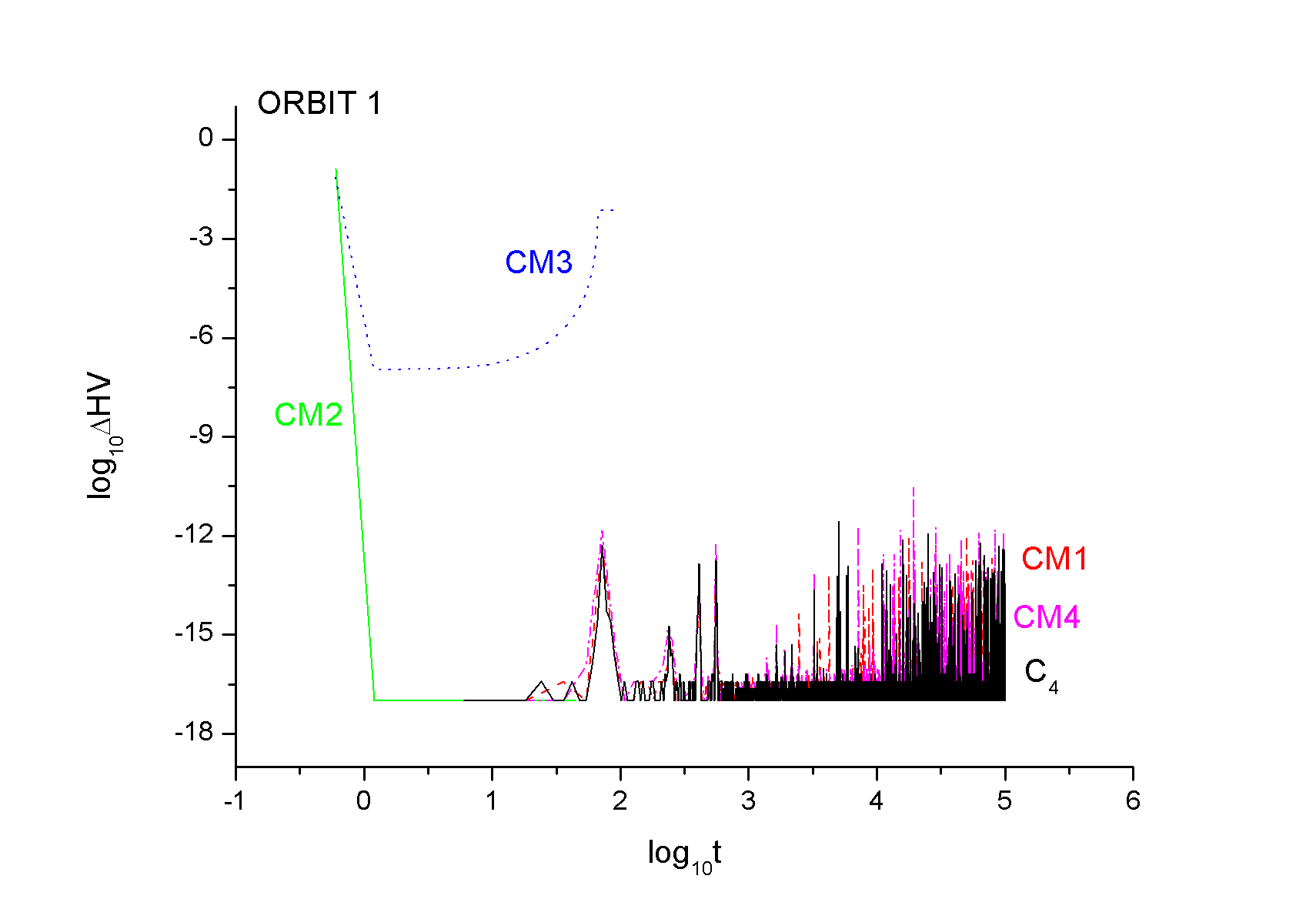}
\caption{Error behaviors of potential energy $V$, $\Delta HV=|V(\mathbf{r}^*)-\frac{\widetilde{V}(\mathbf{r},\widetilde{\mathbf{r}})}{2}|$. Here $\mathbf{r}^*$ denotes the position after corrections, while $\mathbf{r}$ before. The error of $\textbf{C}_4$ (black), $\textbf{CM1}$ (red) and $\textbf{CM4}$ (purple) do not differ much in either magnitude or stability. $\textbf{CM2}$ (green) and $\textbf{CM3}$ (blue) show much larger errors and cause interruptions during the calculations.}
\label{fig4}
\end{figure}
\begin{figure}
\centering
\includegraphics[width=0.4\textwidth]{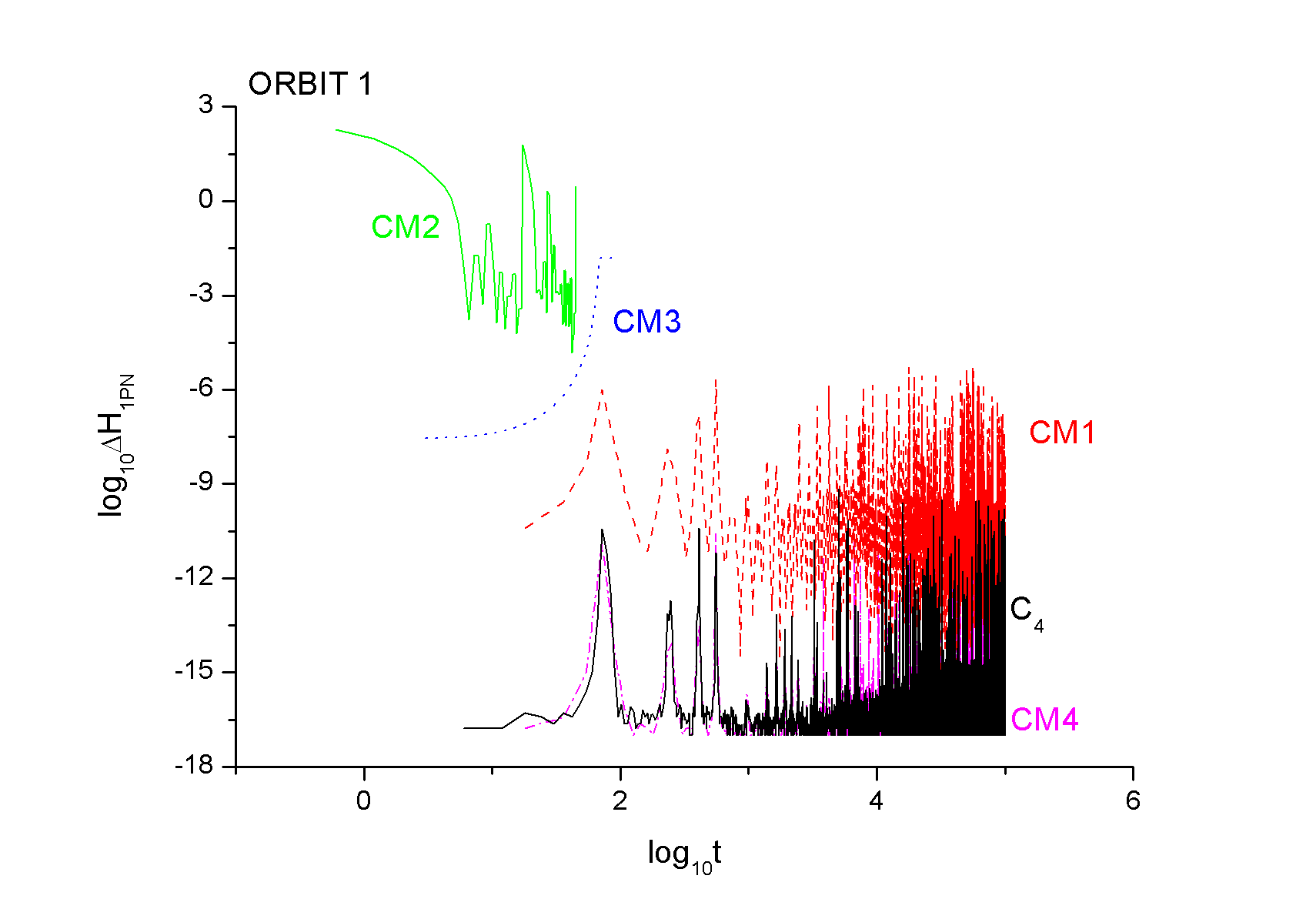}
\caption{Error behaviors of of $H_{1pn}$, $\Delta H_{1pn}=|H_{1pn}(\mathbf{r}^*,\mathbf{p}^*)-\frac{\widetilde{H}_{1pn}(\mathbf{r},\widetilde{\mathbf{r}},\mathbf{p},\widetilde{\mathbf{p}})}{2}|$. $\textbf{C}_4$ (black) and $\textbf{CM4}$ (purple) behave similarly with minimum biases in $H_{1pn}$, where $\textbf{CM2}$ (green) and $\textbf{CM3}$ (blue) show large errors. The bias of $\textbf{CM1}$ (red) is between $\textbf{CM4}$ and $\textbf{CM2}$.}
\label{fig5}
\end{figure}
\begin{figure}
\centering
\includegraphics[width=0.4\textwidth]{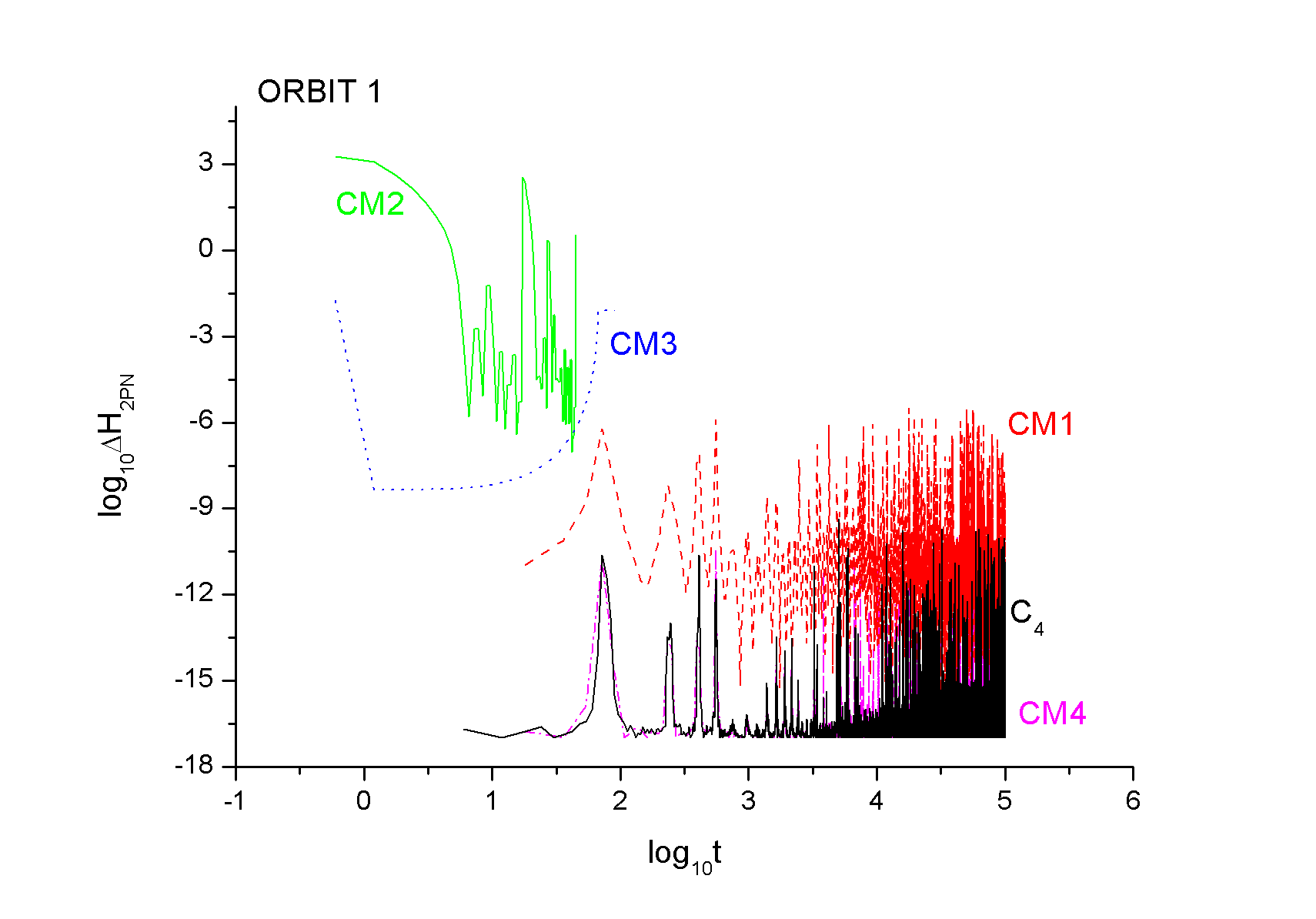}
\caption{Error behaviors of of $H_{2pn}$, $\Delta H_{2pn}=|H_{2pn}(\mathbf{r}^*,\mathbf{p}^*)-\frac{\widetilde{H}_{2pn}(\mathbf{r},\widetilde{\mathbf{r}},\mathbf{p},\widetilde{\mathbf{p}})}{2}|$. The error performance of $\textbf{C}_4$ (black), $\textbf{CM1}$ (red), $\textbf{CM2}$ (green), $\textbf{CM3}$ (blue) and $\textbf{CM4}$ (purple) in $H_{2pn}$ is the same as that in $H_{1pn}$.}
\label{fig6}
\end{figure}
\begin{figure}
\centering
\includegraphics[width=0.4\textwidth]{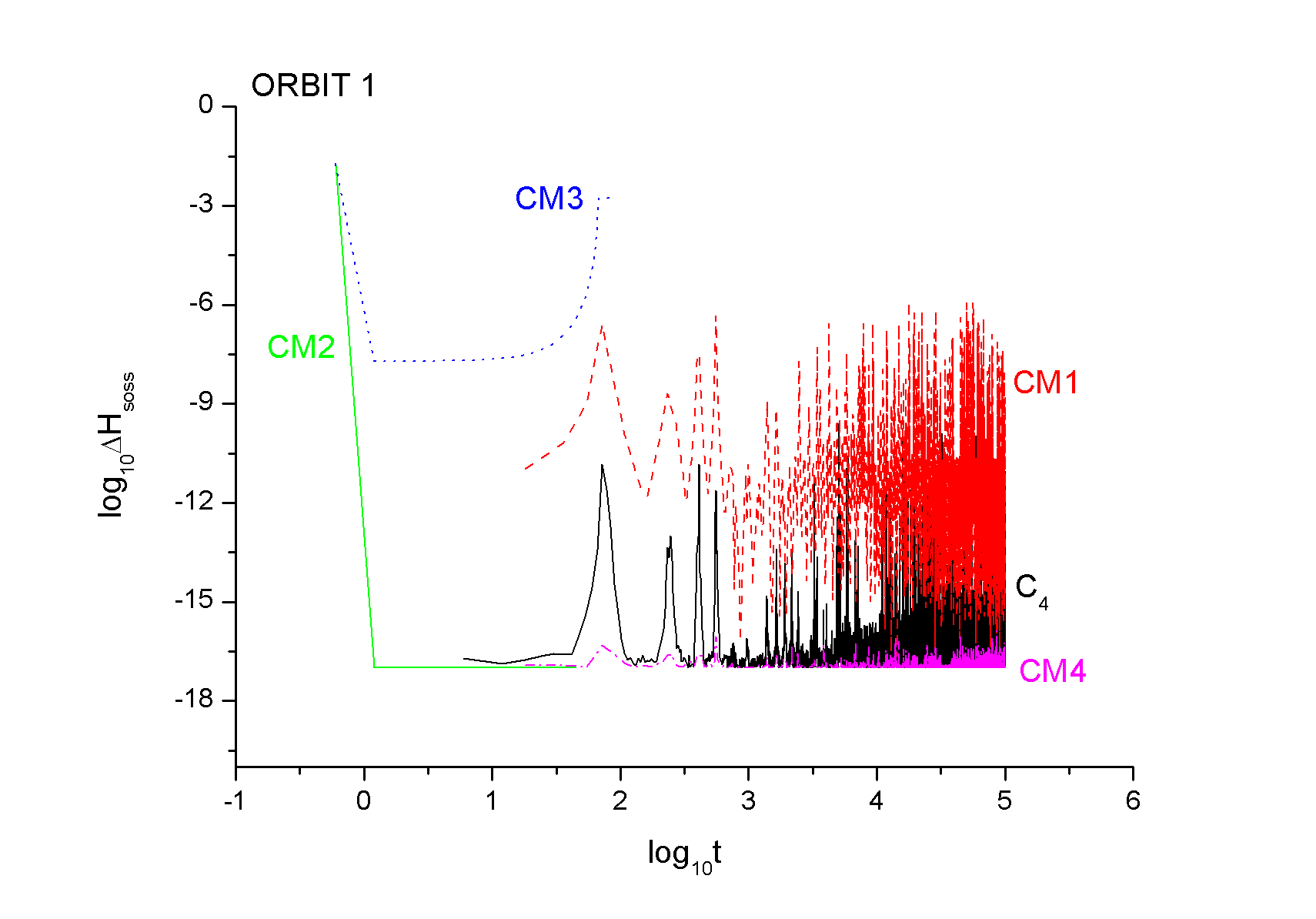}
\caption{Error behaviors of $H_{SOSS}$, $\Delta (H_{SOSS})=|H_{SOSS}(\mathbf{R}^*,\mathbf{P}^*)-\frac{\widetilde{H}_{SOSS}(\mathbf{R},\mathbf{P})}{2}|$. The energy bias of $\textbf{CM4}$ (purple) is minimum, reaching the limit of computer double precision. The error of $\textbf{C}_4$ (black) is slightly larger and increases slowly. The long-term error of $\textbf{CM1}$ (red) is stable but larger than $\textbf{C}_4$ and $\textbf{CM4}$. $\textbf{CM2}$ (green) and $\textbf{CM3}$ (blue) have large errors.}
\label{fig7}
\end{figure}

However, more detailed differences will be revealed when we compare the changes of each subterm in Hamiltonian $\widetilde{H}$ before and after the corrections. Fig. \ref{fig3} shows the energy error of the kinetic energy term. $\textbf{CM4}$ gets the minimum bias in the kinetic energy term, while the errors of $\textbf{CM2}$ and $\textbf{CM3}$ increase sharply. The errors of the potential energy term are drawn in Fig. \ref{fig4}, where $\textbf{C}_4$, $\textbf{CM1}$, and $\textbf{CM4}$ have similar error behavior, while $\textbf{CM2}$ and $\textbf{CM3}$ perform poorly. The errors of $H_{1PN}$ and $H_{2PN}$ are drawn in Fig. \ref{fig5} and Fig. \ref{fig6}, respectively. These two figures look similar to each other. $\textbf{CM4}$ and $\textbf{C}_4$ have the minimum biases. $\textbf{CM1}$ is also stable but with larger bias. $\textbf{CM2}$ and $\textbf{CM3}$ produce more obvious subterm energy errors. The errors of rest subterms, i.e., $H_{SOSS}$, are also drawn in Fig. \ref{fig7}, where $\textbf{CM4}$ behaves well with minimum biases. The map matrix of $\textbf{CM1}$ biases the values of $H_{SOSS}$ significantly, while $\textbf{CM2}$ and $\textbf{CM3}$ fail to restrain the bias.

\begin{figure}
\centering
\includegraphics[width=0.4\textwidth]{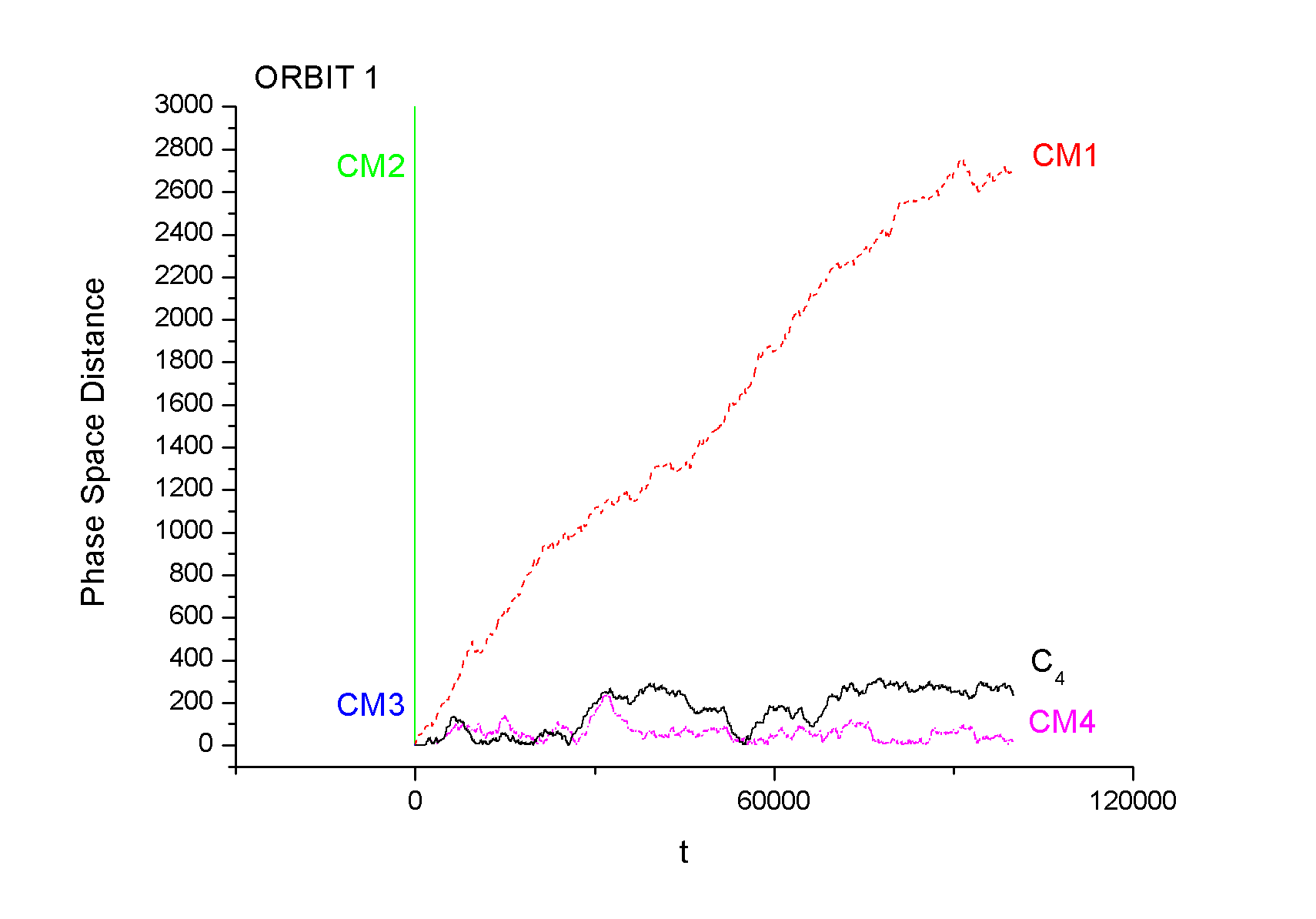}
\caption{Phase space distance $D$ between $8(9)RKF$ and other algorithms as functions of time steps. The distances in the ascending order are $\textbf{CM4}$ (purple), $\textbf{C}_4$ (black), $\textbf{CM1}$ (red). 
The $D$ of $\textbf{CM2}$ (green) and $\textbf{CM3}$ (blue) are very large at the beginning, indicating that their numerical solutions are not available.}
\label{fig8}
\end{figure}

What is more, we can conclude that the matrix $M4$ of $\textbf{CM4}$ in a degree does not change the values of the subterms of the Hamiltonian, while the $\textbf{C}_4$ subterms experience a certain degree of energy exchange between each other. In order to ensure total energy conservation, $\textbf{CM1}$ undergoes significant subterm energy biases. $\textbf{CM2}$ and $\textbf{CM3}$ ensure both energy and angular momentum conversations, and the biases in energy of each subterm are large. So which algorithm is closer to the real solution? We compare the phase-space distance $D$ of $\textbf{CM1}$, $\textbf{C}_4$, $\textbf{CM2}$, $\textbf{CM3}$ and $\textbf{CM4}$ with respect to that of $8(9)RKF$ at every integration step in Fig. \ref{fig8} to know which correction map is more accurate.
Here $D=\sqrt{(\textbf{R}_{RK}-\textbf{R})^{2}+(\textbf{P}_{RK}-\textbf{P})^{2}}$, the solutions of $8(9)RKF$ are denoted as $\textbf{R}_{RK}$ and $\textbf{P}_{RK}$.  We find that $D$ of $\textbf{CM4}$, which reduces the energy bias of each subterm, is minimum. $\textbf{C}_4$, which keeps the original Hamiltonian $H$ equaling to half of the new Hamiltonian $\widetilde{H}$, ranks the second. $D$ of $\textbf{CM1}$ is larger than that of $\textbf{C}_4$ and increases quickly with time steps. $\textbf{CM2}$ and $\textbf{CM3}$, which ensure both energy and angular momentum conservation, show a very dramatic growth of $D$.

Generally, most of the manifold correction methods adjust the momenta or positions to preserve the conserved quantities \citep{Wu_2007APR, Ma:2008ApJ, Wang_2018AAS}. They suppress the accumulated errors in most cases and bring the numerical solution closer to the exact solution. However, from the relative energy error in Fig. \ref{fig2} and the phase space distance in Fig. \ref{fig8}, it can be seen that $\textbf{CM1}$, $\textbf{CM2}$ and $\textbf{CM3}$ make their numerical solutions away from the exact ones, although they try to minimize the errors in the constants of motion. 

\begin{figure}
\centering
\includegraphics[width=0.4\textwidth]{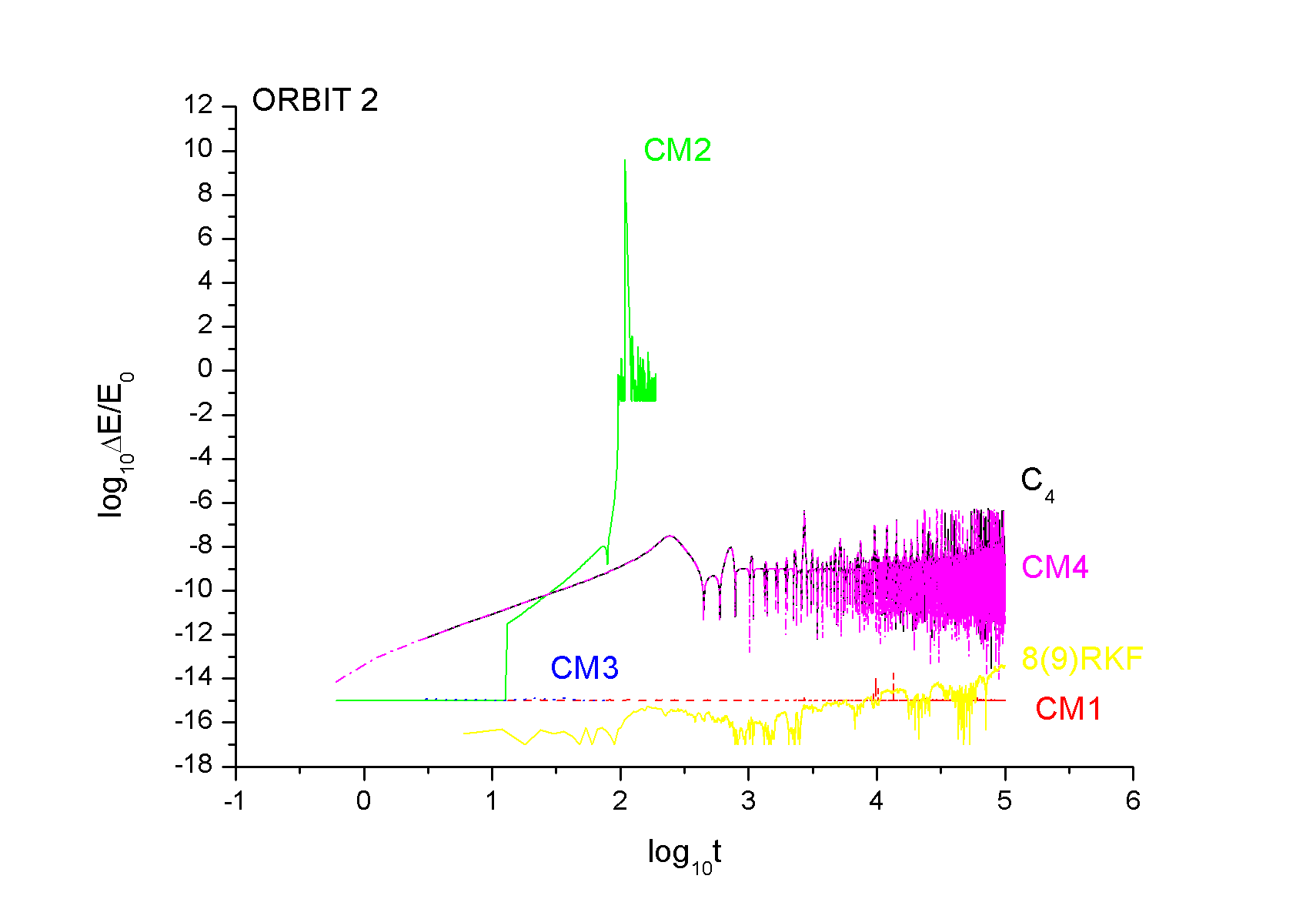}
\caption{Relative Energy error of $H$ for $\textbf{C}_4$ (black), $\textbf{CM1}$ (red), $\textbf{CM2}$ (green), $\textbf{CM3}$ (blue), $\textbf{CM4}$ (purple), $8(9)RKF$ (yellow) in orbit 2, $\Delta E/E_0=|\frac{H(t)-H(0)}{H(0)}|$, where $H(t)$ is the Hamiltonian $H$ at time t, while $H(0)$ represents the initial energy.}
\label{fig9}
\end{figure}

\begin{figure}
\centering
\includegraphics[width=0.4\textwidth]{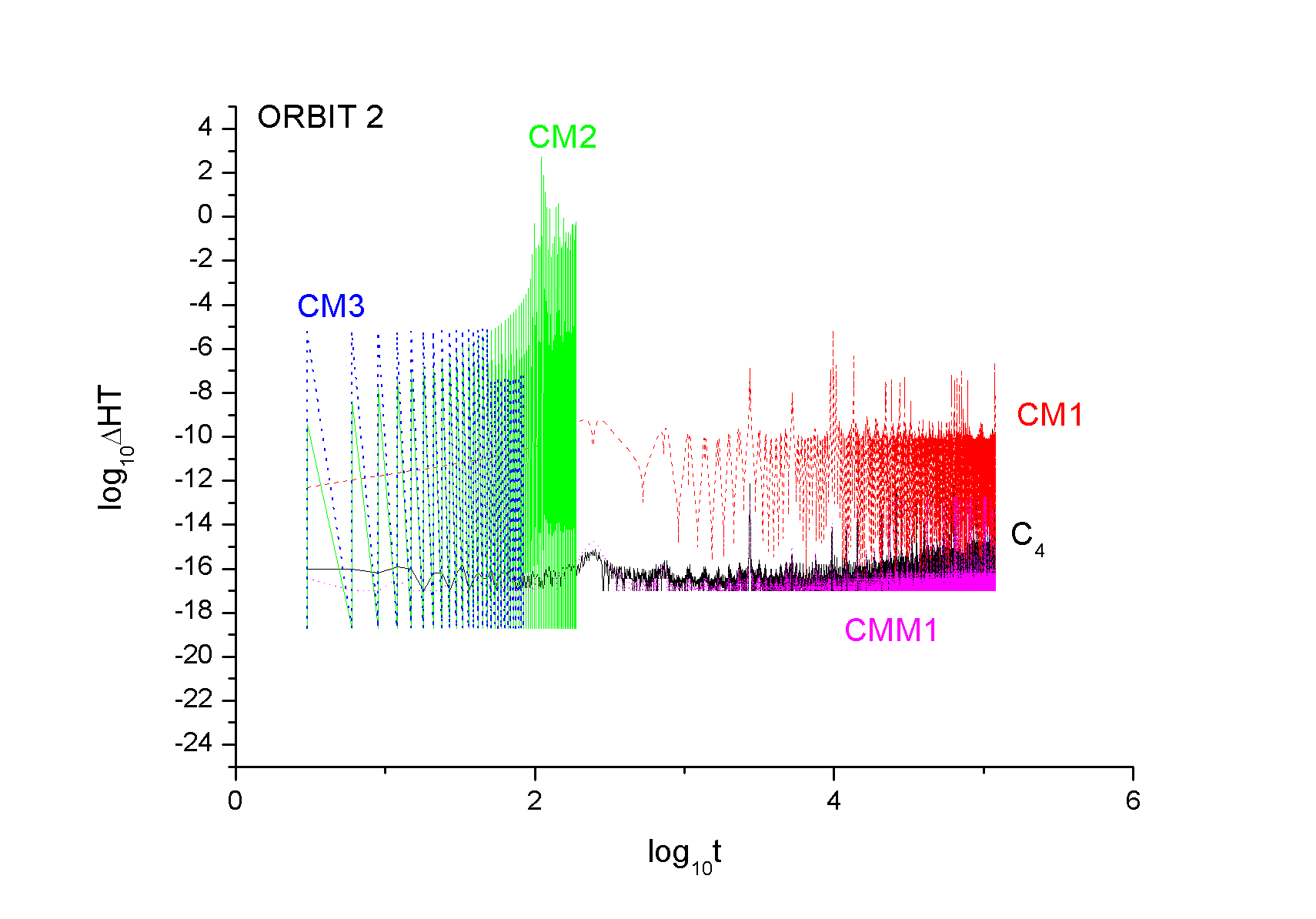}
\caption{Error behaviors of kinetic energy $T$ for $\textbf{C}_4$ (black), $\textbf{CM1}$ (red), $\textbf{CM2}$ (green), $\textbf{CM3}$ (blue), $\textbf{CM4}$ (purple) in orbit 2, $\Delta HT=|T(\mathbf{p}^*)-\frac{\widetilde{T}(\mathbf{p},\widetilde{\mathbf{p}})}{2}|$. Here $\mathbf{p}^*$ denotes the momentum after corrections, while $\mathbf{p}$ before.}
\label{fig10}
\end{figure}

\begin{figure}
\centering
\includegraphics[width=0.4\textwidth]{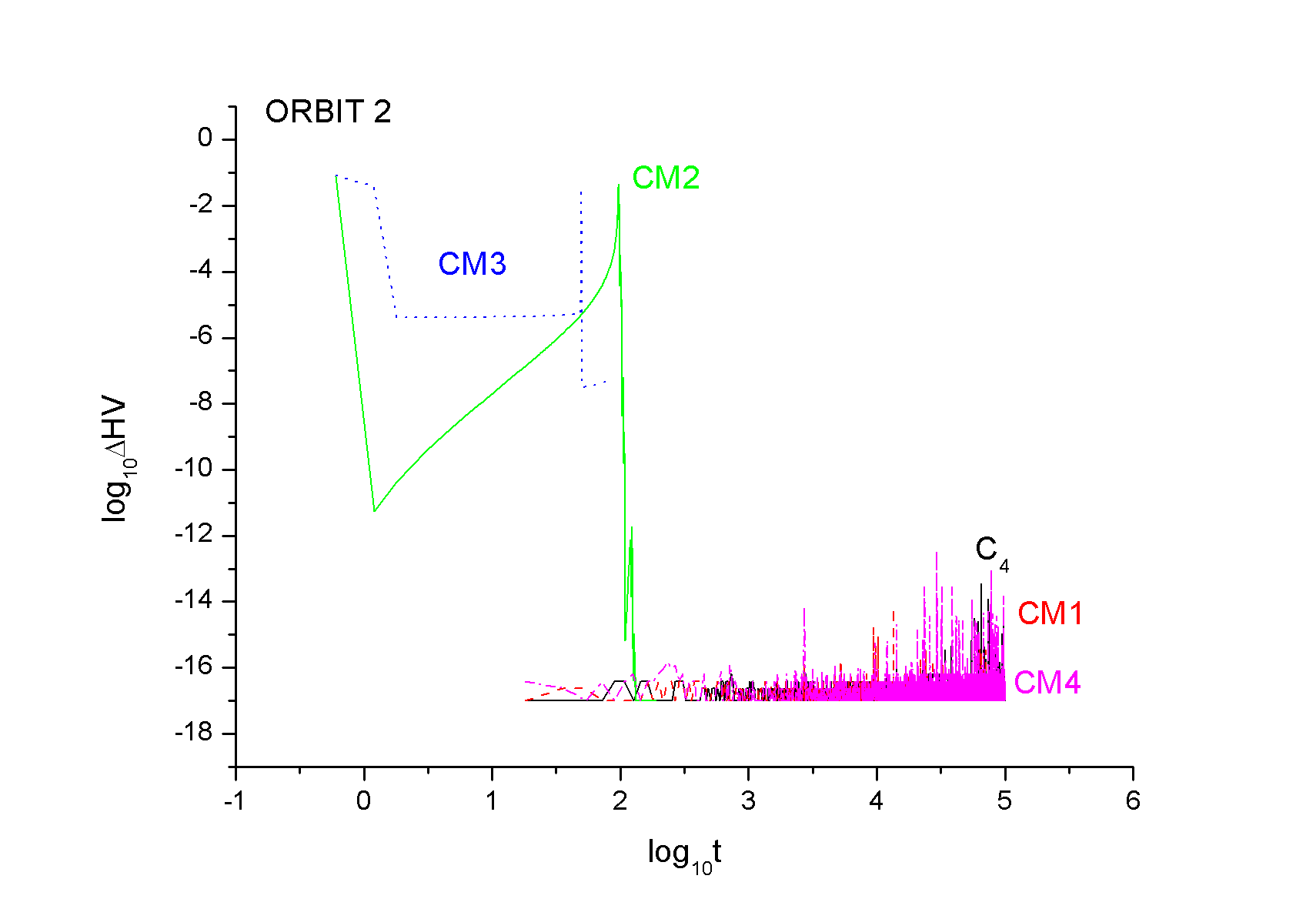}
\caption{Error behaviors of potential energy $V$ for $\textbf{C}_4$ (black), $\textbf{CM1}$ (red), $\textbf{CM2}$ (green), $\textbf{CM3}$(blue), $\textbf{CM4}$ (purple) in orbit 2, $\Delta HV=|V(\mathbf{r}^*)-\frac{\widetilde{V}(\mathbf{r},\widetilde{\mathbf{r}})}{2}|$. Here $\mathbf{r}^*$ denotes the position after correction, while $\mathbf{r}$ before.}
\label{fig11}
\end{figure}

\begin{figure}
\centering
\includegraphics[width=0.4\textwidth]{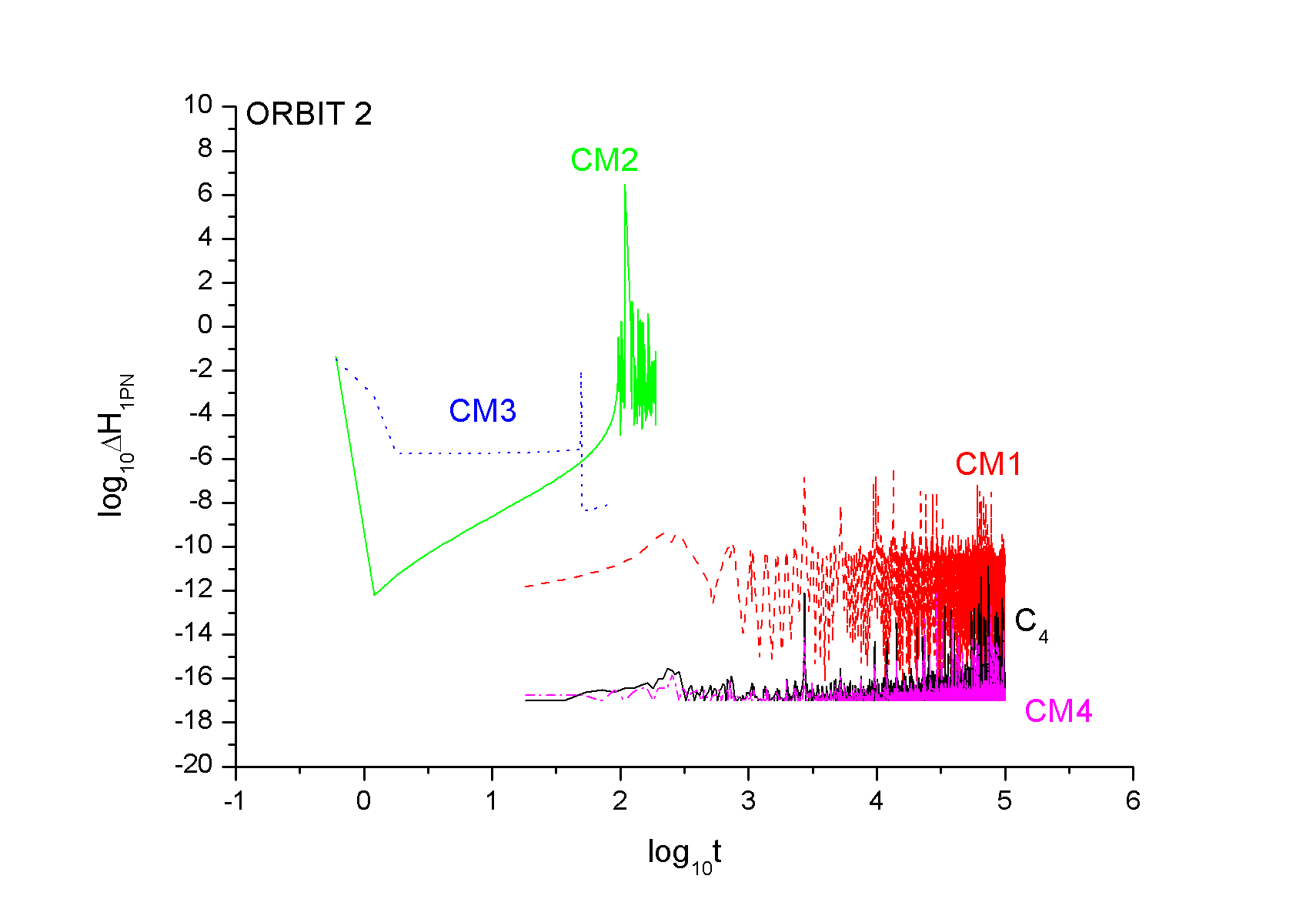}
\caption{Error behaviors of $H_{1pn}$ for $\textbf{C}_4$ (black), $\textbf{CM1}$ (red), $\textbf{CM2}$ (green), $\textbf{CM3}$ (blue), $\textbf{CM4}$ (purple) in orbit 2, $\Delta H_{1pn}=|H_{1pn}(\mathbf{r}^*,\mathbf{p}^*)-\frac{\widetilde{H}_{1pn}(\mathbf{r},\widetilde{\mathbf{r}},\mathbf{p},\widetilde{\mathbf{p}})}{2}|$}
\label{fig12}
\end{figure}

\begin{figure}
\centering
\includegraphics[width=0.4\textwidth]{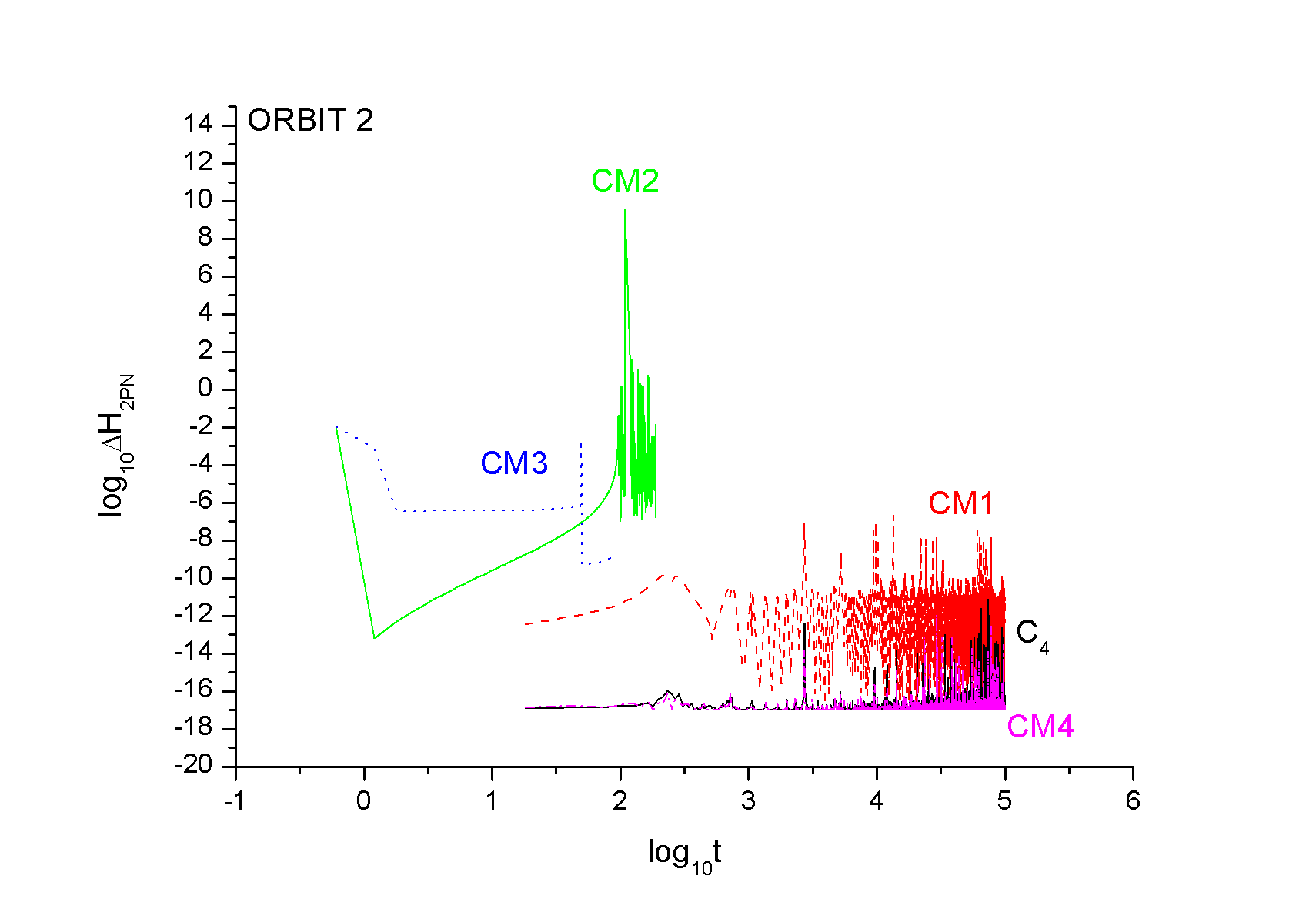}
\caption{Error behaviors of of $H_{2pn}$ for $\textbf{C}_4$ (black), $\textbf{CM1}$ (red), $\textbf{CM2}$ (green), $\textbf{CM3}$ (blue), $\textbf{CM4}$ (purple) in orbit 2, $\Delta H_{2pn}=|H_{2pn}(\mathbf{r}^*,\mathbf{p}^*)-\frac{\widetilde{H}_{1pn}(\mathbf{r},\widetilde{\mathbf{r}},\mathbf{p},\widetilde{\mathbf{p}})}{2}|$.}
\label{fig13}
\end{figure}

\begin{figure}
\centering
\includegraphics[width=0.4\textwidth]{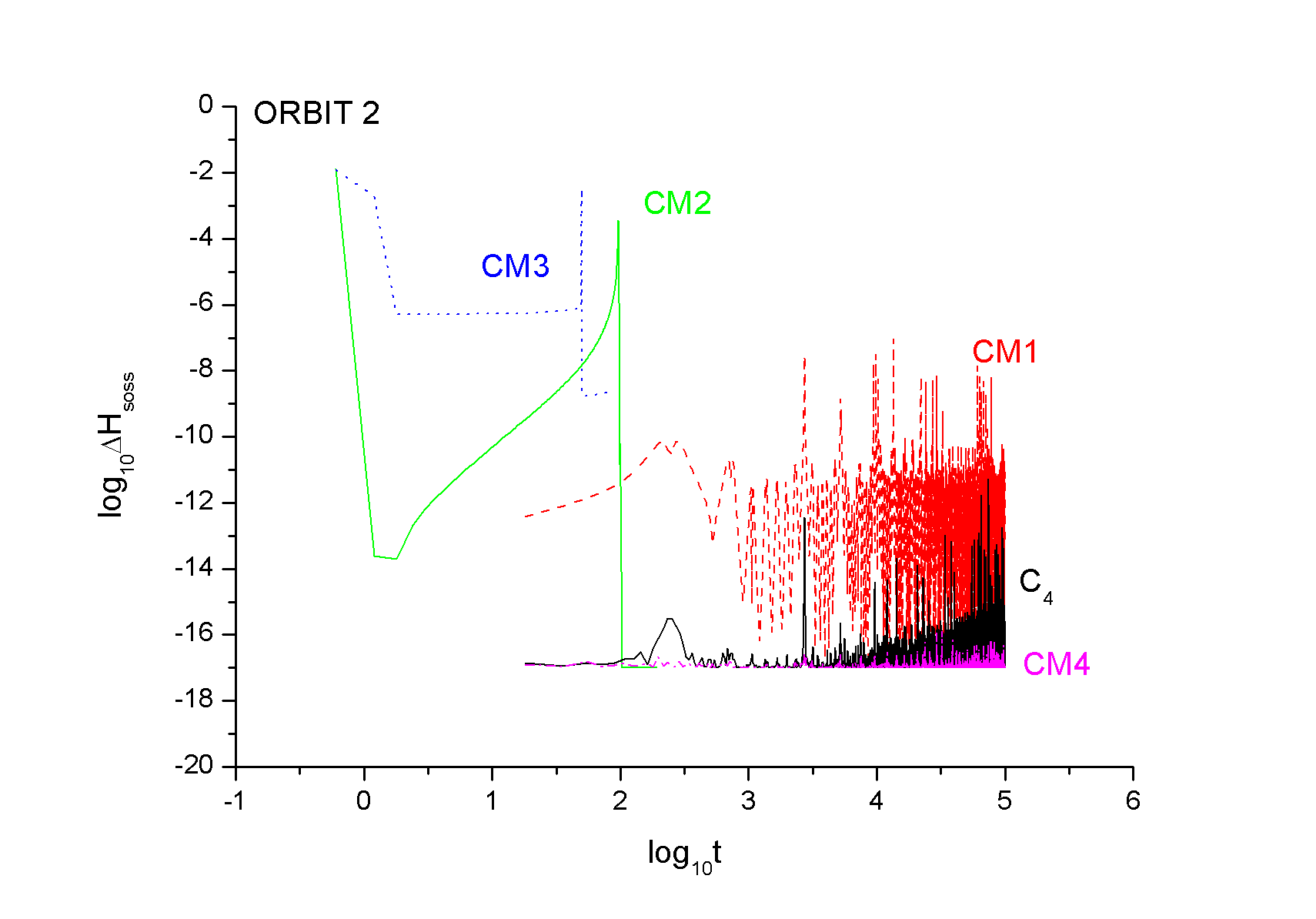}
\caption{Error behaviors of $H_{1.5so}+H_{2ss}$ for $\textbf{C}_4$(black), $\textbf{CM1}$(red dash), $\textbf{CM2}$(green), $\textbf{CM3}$(blue dot), $\textbf{CM4}$(Purple dash dot) in orbit 2, $\Delta (H_{1.5so}+H_{2ss})=H_{1.5so}(\mathbf{R}_a+\mathbf{P}_a)+H_{2ss}(\mathbf{R}_a+\mathbf{P}_a)-\frac{\widetilde{H}_{1.5so}+\widetilde{H}_{2ss}}{2}$.}
\label{fig14}
\end{figure}

To consolidate our conclusion, we perform numerical simulations for another orbit, called orbit 2, with initial conditions $(\beta;\textbf{r},\textbf{p})=(1;8.309,0,0,0,0.5,0),\chi_{1}=\chi_{2}=1,\hat{\textbf{S}}_{1}=(0.13036,0.262852,-0.983734),\hat{\textbf{S}}_{2}=(0.118966,-0.13459,-0.983734)$. We set $w_1=100$, $w_2=1$ and $h=0.6$. We plot $\Delta E/E_0$, $\Delta T$, $\Delta V$, $\Delta H_{1PN}$, $\Delta H_{2PN}$ and $\Delta H_{SOSS}$ in Fig. \ref{fig9}, \ref{fig10}, \ref{fig11}, \ref{fig12}, \ref{fig13} and \ref{fig14} respectively. The performance of each algorithm in orbit 2 is not very different from that in orbit 1. The energy change of each subterm after the application of the $\textbf{CM4}$'s map is minimal, and its numerical solution is the closest to that of the $8(9)RKF$. On the contrary, although $\textbf{CM1}$, $\textbf{CM2}$, and $\textbf{CM3}$ try to minimize the energy error, their energy biases of most subterms are larger than those in $\textbf{CM4}$. Especially, $\textbf{CM2}$ and $\textbf{CM3}$ develop non-physical evolution, which are far from the solutions calculated by $8(9)RKF$. Finally, we show the time consumption of each algorithm in Table \ref{Table 2}. It can be inferred that although $\textbf{CM4}$ has two more scale factors than $\textbf{C}_4$, the computational difficulty of each correction factor is also much less than $\textbf{C}_4$. Therefore, $\textbf{C}_4$ and $\textbf{CM4}$ are also close in terms of computational efficiency.
\begin{figure}
\centering
\includegraphics[width=0.4\textwidth]{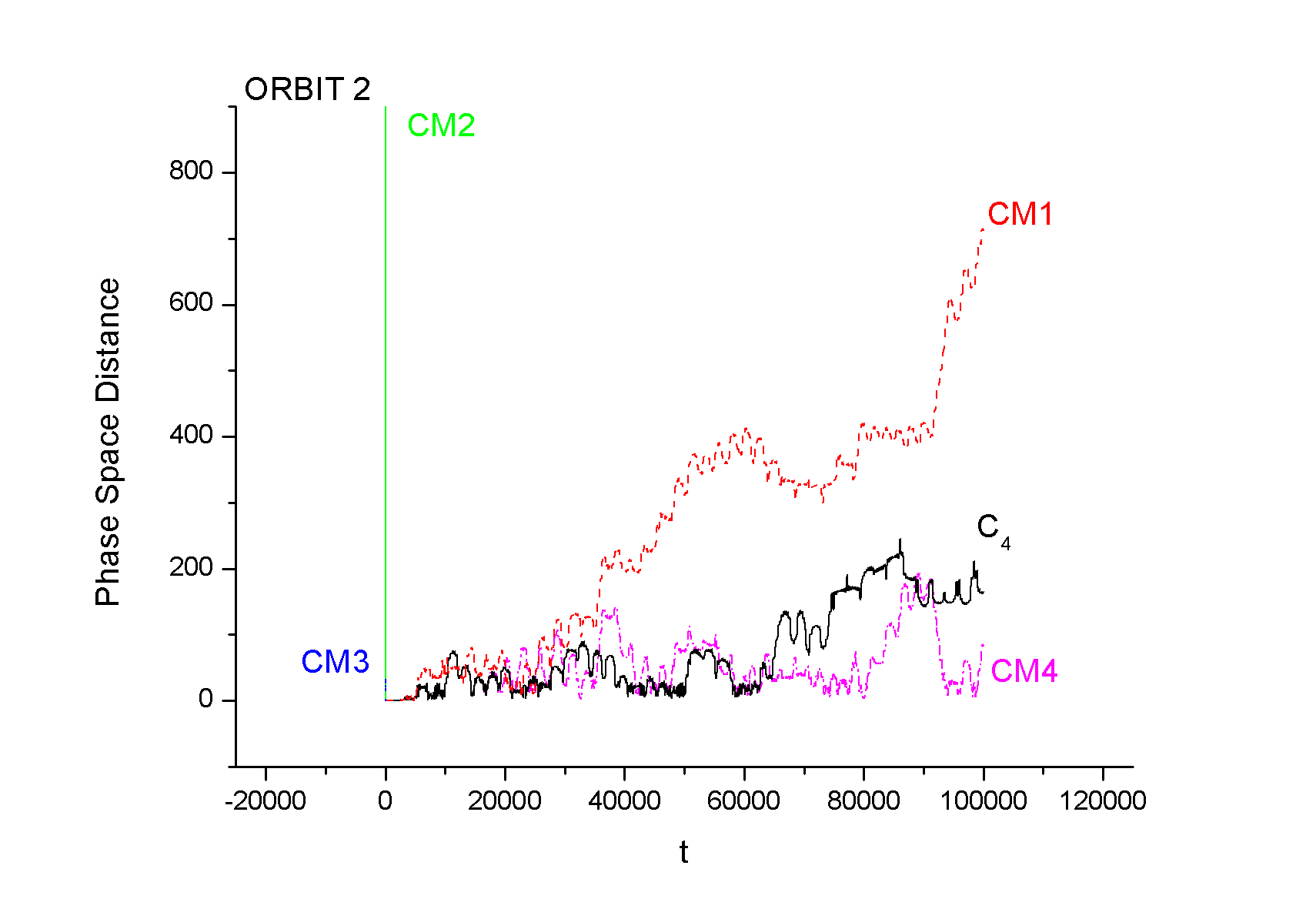}
\caption{Phase-space distance with $8(9)RKF$ for other algorithms at every corresponding integration steps. The distances to the exact solution in descending order are $\textbf{CM4}$(Purple dash dot), $\textbf{C}_4$(black), $\textbf{CM1}$(red dash). $\textbf{CM2}$(green) and $\textbf{CM3}$(blue dot), although the final distances are not known without completing the calculation, are already very exaggerated from the beginning, indicating that their numerical solutions are not available.}
\label{fig15}
\end{figure}

\begin{table}
\begin{center}
\small \caption{CPU times (hour: minute: second) for the four algorithms calculating the orbit 1 and orbit2 of spinning compact binaries from 0 to $10^5$ integration step.}
\label{Table 2}
\begin{tabular}{cccccc}\hline
methods & $C_4$ & $\textbf{CM1}$ & $\textbf{CM2}$/$\textbf{CM3}$ & $\textbf{CM4}$ & 8(9)RKF\\
\hline
 orbit1 & 0:0:25 & 0:0:26 & unknown & 0:0:28 & 0:0:53\\
 orbit2 & 0:0:25 & 0:0:25 & unknown & 0:0:27 & 0:0:50\\
\hline
\end{tabular}
\end{center}
\end{table}
\section{Summary}\label{sec:4}
The spinning compact binaries system is one of the gravitational-wave sources for broadband laser interferometers. In order to analyze the gravitational wave signals using matched filtering techniques,
reliable numerical algorithms are needed. Spinning compact binaries in PN celestial mechanics have inseparable Hamiltonian. The extended phase-space algorithm is an effective solution for the problem of this system.  

In this paper, we test the extended phase-space algorithms with different correction maps in the chaotic orbits of spinning compact binaries. $\textbf{C}_4$ ensures that the original Hamiltonian $H(\textbf{R},\textbf{P})$ is always equal to the half of the new Hamiltonian $\widetilde{H}(\textbf{R},\widetilde{\textbf{R}},\textbf{P},\widetilde{\textbf{P}})$, but leads to some energy biases of the subterms of the Hamiltonian. $\textbf{CM1}$ minimizes the energy error. After its correction, the numerical solutions show significant biases in the subterms of energy. CM2 takes the total energy and total angular momentum conservation into account. $\textbf{CM3}$ uses the least-squares correction and adds a correction for the spin length, but its performance is close to that of $\textbf{CM2}$ and shut down during calculation. $\textbf{CM4}$ is designed to ensure that the subterms of the original Hamiltonian is equal to the half of that of the new Hamiltonian. $\textbf{CM1}$, $\textbf{CM2}$ and $\textbf{CM3}$ keep the constants of motion but make the solution more away from the exact one after correction. When the manifold correction method is applied to the Runge-Kutta algorithm, the suppression of the energy error growth may bring more benefits to the calculation, but the extended phase-space algorithm without any map is an explicit symplectic algorithm for the new Hamiltonian $\widetilde{H}$, which is already very accurate for the total energy and each subterm energy itself, and forcing the integrated solution back to the original hypersurface will lead to inaccurate energy calculation of each subterm and away from the exact solution. Numerical simulations are not only for the sake of reducing total energy biases, but also for getting the results closer to physical realities. Besides, the CPU time is also an important factor for choosing one algorithm. To this end, we do not recommend $\textbf{CM1}$, $\textbf{CM2}$ and $\textbf{CM3}$, but suggest $\textbf{CM4}$ to calculate the chaotic orbits of spinning compact binaries.

\section*{Acknowledgements}
This work is supported in part by the National Natural Science Foundation of China (NSFC) under Grant Nos. 12203108, 11875327, 12275367 and 12073089, the Fundamental Research Funds for the Central Universities, and the Sun Yat-sen University Science Foundation.
\section*{Data availability}
The data underlying this article are available in the article and in its online supplementary material.




\bibliographystyle{mnras}
\bibliography{ref} 

\label{lastpage}
\end{document}